\documentclass[12pt]{article}
\pdfoutput=1
\usepackage{jheppub}
\usepackage[utf8]{inputenc}
\usepackage{graphicx}
\usepackage{amsfonts}
\usepackage[dvipsnames]{xcolor}
\usepackage{enumerate}
\usepackage{amsmath}
\usepackage{amssymb}
\usepackage[mathscr]{eucal}
\usepackage{slashed}
\usepackage{array}
\usepackage{amsthm}
\usepackage{epsfig}
\usepackage{wasysym} 
\usepackage{tikz-cd} 
\usepgfmodule{shapes}
\usetikzlibrary{backgrounds, arrows,calc,shapes,decorations.pathreplacing, decorations.markings, automata,positioning}
\usepackage{ytableau} 
\usepackage{dirtytalk} 
\usepackage{cancel} 
\usepackage{changepage} 
\usepackage{float} 

\tikzset{
  arrow/.pic={\path[tips,every arrow/.try,->,>=#1] (0,0) -- +(.1pt,0);},
  pics/arrow/.default={latex,very thick}
}

\newcommand\bovermat[2]{%
\makebox[0pt][l]{$\smash{\overbrace{\phantom{%
\begin{matrix}#2\end{matrix}}}^{\text{#1}}}$}#2}

\newcommand{\be}{\begin{equation}}
\newcommand{\ee}{\end{equation}}
\newcommand{\bea}{\begin{eqnarray}}
\newcommand{\eea}{\end{eqnarray}}
\newcommand{\ba}{\begin{array}}
\newcommand{\ea}{\end{array}}
\newcommand{\bpic}{\begin{tikzpicture}}
\newcommand{\epic}{\end{tikzpicture}}

\newcommand{\tr}{\text{tr}\,}
\newcommand{\Pf}{\text{Pfaff}\,}

\newcommand{\M}{{\mathfrak M}}

\newcommand{\llra}{\longleftrightarrow}

\newcommand\qt{\tilde q}

\newcommand\bt{\tilde b}

\newcommand\Qt{\tilde Q}

\newcommand{\CT}{{\mathcal T}}

\newcommand{\CW}{{\mathcal W}}

\newcommand{\cN}{\mathcal{N}}
\newcommand{\cO}{\mathcal{O}}

\newcommand{\cR}{\mathcal{R}}

\newcommand{\cT}{\mathcal{T}}

\newcommand{\cW}{\mathcal{W}}

\def\a{\alpha} 
\def\b{\beta}

\def\e{\varepsilon}

\def\s{\sigma}

\def\P{\Phi}

\title{Sequential deconfinement and self-dualities in $4d$ $\mathcal{N}\!=\!1$ gauge theories}

\author[1]{Stephane Bajeot}
\author[2]{Sergio Benvenuti}

\affiliation[1]{SISSA, Via Bonomea 265, 34136 Trieste, Italy}
\affiliation[2]{INFN, Sezione di Trieste, Via Valerio 2, 34127 Trieste, Italy}
\emailAdd{sbajeot@sissa.it, benve79@gmail.com}

\abstract{We apply the technique of sequential deconfinement to the four dimensional $\mathcal{N}\!=\!1$ $Usp(2N)$ gauge theory with an antisymmetric field and $2F$ fundamentals. The fully deconfined frame is a length-$N$ quiver. We use this deconfined frame to prove the known self-duality of $Usp(2N)$ with an antisymmetric field and $8$ fundamentals. 
Along the way we encounter a subtlety: in certain quivers with degenerate holomorphic operators, a naive application of Seiberg duality rules leads to an incorrect superpotential or chiral ring.

We also consider the reduction to $3d$ $\mathcal{N}\!=\!2$ theories, recovering known fully deconfined duals of $Usp(2N)$ and $U(N)$ gauge theories, and obtaining new ones.}

\begin{document}

\maketitle

\section{Introduction and summary}
Dualities are among the most powerful tool to analyze quantum field theories at strong coupling. In this paper we are interested in four dimensional $\cN=1$ gauge theories with rank-$2$ matter.
Various recent works derived dualities involving gauge theories with rank-$2$ matter using only basic dualities involving gauge theories with fundamental matter, like Seiberg \cite{Seiberg:1994pq}, Intriligator-Pouliot \cite{Intriligator:1995ne} and Aharony dualities \cite{Aharony:1997gp}.

In three dimensions, \cite{Pasquetti:2019uop,Pasquetti:2019tix} proved a $3d$ $\cN=2$ S-confining duality for $U(N)$ with adjoint and $(1,1)$ fundamentals \cite{Aghaei:2017xqe, Benvenuti:2018bav, Amariti:2018wht} iterating Aharony duality. \cite{Benvenuti:2020gvy} proved $3d$ $\cN=2$ self-dualities of $U(N)$ with adjoint and $(2,2)$ fundamentals and of $Usp(2N)$ with antisymmetric and $6$ fundamentals \cite{Benvenuti:2018bav, Amariti:2018wht}, iterating Aharony dualities. Iterative application of confining monopole dualities \cite{Benini:2017dud} have been proven very useful in $3d$ also in \cite{Benvenuti:2017lle, Benvenuti:2017kud, Benvenuti:2017bpg, Giacomelli:2017vgk}.

More recently, in four dimensions, \cite{Bajeot:2022kwt, Bottini:2022vpy} proved the S-confinement of  $4d$ $\cN=1$ $Usp(2N)$ with antisymmetric and $6$ fundamentals, and \cite{Bajeot:2022kwt} proved that all S-confining quivers with one node (classified long time ago in \cite{Csaki:1996sm, Csaki:1996zb}) can be obtained from Seiberg and Intriligator-Pouliot S-confinements.

By \emph{proving the duality} between theory $\cT_1$ and theory $\cT_k$, we mean constructing a sequence of quiver gauge theories $\cT_i$, $i=1,\ldots, k$, such that $\cT_i$ is related to $\cT_{i+1}$ by the application of an elementary duality on a single node. So all the theories $\cT_1,\ldots,\cT_k$ are infrared dual. Assuming, as is standard, that the renormalization group flows commute with dualizing a single node, this amounts to a proof of the non-elementary duality $\cT_1 \leftrightarrow \cT_k$. The strategy to construct the sequence of dual quivers requires the \emph{deconfinement} of some rank-$2$ matter. The deconfinement we need in this paper was introduced in \cite{Bajeot:2022kwt, Bottini:2022vpy}, and uplifts the $3d$ deconfiments used in \cite{Pasquetti:2019uop,Pasquetti:2019tix, Benvenuti:2020gvy}.\footnote{Similar deconfinements appear in \cite{Garcia-Etxebarria:2012ypj, Garcia-Etxebarria:2013tba} in the study of orientifolded dimer models, and were used in \cite{Etxebarria:2021lmq} to construct $\cN=1$ Lagrangians for $4d$ $\cN=2$ SCFTs.}

In a related context, \cite{Bottini:2021vms, Hwang:2021ulb} proved the $4d$ mirrors dualities \cite{Hwang:2020wpd} (upflit of $3d$ $\cN=4$ mirror symmetries \cite{Intriligator:1996ex, Hanany:1996ie} for linear quivers) using only Intriligator-Pouliot dualities. One difference is that, in the mirror symmetry case, the intermediate steps are quasi-Lagrangian theories, that is they involve gauging symmetries which only emerge in the infrared, as in \cite{Gadde:2015xta}.

In this paper we work out the \emph{sequentially deconfined} dual of $4d$ $\cN=1$ $Usp(2N)$ with antisymmetric and $2F$ fundamentals, uplifting the $3d$ $\cN=2$ results of  \cite{Benvenuti:2020gvy}. This means that we step by step prove a duality with a linear quiver gauge theories with $N$ nodes and a certain \emph{saw} structure, full details of the theory are in \eqref{stepFinalSp}. This fully deconfined dual frame enjoys the nice property that all the chiral ring operators are gauge singlet fields, similarly to what happens for Intriligator-Pouliot and Aharony dualities.\footnote{This property cannot be enjoyed by theories with general baryonic operators in the chiral ring, such as in $SU(N)$ SQCD or in the sequentially deconfined $3d$ $\cN=2$ $SO$ gauge theories \cite{BenvenutiLoMonaco}.}

As an application of this construction, we provide a simple proof of the \emph{self-duality modulo flips}\footnote{Flipping a gauge invariant operator $\cO$ means adding a gauge singlet field $\sigma$ to the theory, and an interaction term of the form $\s \cO$ to the superpotential. The gauge singlet $\s$ is called a \emph{flipper}.} of $Usp(2N)$ with antisymmetric and $8$ fundamentals, proposed long time ago in \cite{Csaki:1997cu}.  By \emph{self-dual modulo flips} we mean that the electric and magnetic theory share the same gauge structure, but differ by gauge singlets fields of flip type. Self-dualities modulo flips have been discussed in \cite{Distler:1996ub, Csaki:1997cu, Karch:1997jp, Razamat:2018gbu}, the simplest case is $SU(2)$ with $8$ doublets. Interestingly, given a self-duality modulo flips, it is possible to move the singlets across the duality and construct \emph{exactly self-dual theories} with enhanced infra-red global symmetry, see for instance \cite{Razamat:2017wsk, Razamat:2018gbu, Sela:2019nqa, Hwang:2020ddr, Hwang:2021xyw}. In our case of $Usp(2N)$ with antisymmetric and $8$ fundamentals, \cite{Hwang:2020ddr} discussed various exactly self-dual theories and discussed the associated symmetry enhancements. Sometimes these symmetry enhancements can be understood compactifying a $6d$ $(1,0)$ SCFT on a Riemann surface. For $Usp(2N)$ with antisymmetric and $8$ fundamentals, \cite{Hwang:2021xyw} related the self-duality and specific symmetry enhancements to a compactification of the rank-$N$ E-string $6d$ SCFT on a $2$-sphere.

We encounter one subtlety (which is also present in the $3d$ cases but was overlooked in \cite{Benvenuti:2020gvy}) during the process that we call \emph{degenerate holomorphic operator ambiguity}. As the name suggests, this phenomenon appears when we reach a frame that contains more than one gauge invariant holomorphic operator  with the same global symmetry quantum numbers  (including $U(1)_R$), but only one combination is a chiral protected operator. If such an operator is flipped by a gauge singlet, only one specific combination appears in the superpotential. In the examples we encounter in this paper, it happens that if we follow the rules of Seiberg duality (as is usually done) we end up with the incorrect result. In some cases we can determine which is the precise combination of operators appearing in the chiral ring (equivalently, the combination that can be flipped) by going in a dual frame and using classical F-terms relations there. Hence, in the original theory with \emph{degenerate holomorphic operator ambiguity}, the ambiguity is resolved by quantum relations. In the case of $F=4$, the precise superpotential is crucial in the proof of the self-duality modulo flips, so this case provides a good consistency check of our procedure.

\subsubsection*{Future directions}

One natural question is wether there is a relation between our sequentially deconfined dual of $Usp(2N)$ with an antisymmetric and $2F$ fundamentals and Kutasov-Schwimmer type dualities \cite{Kutasov:1995ve, Kutasov:1995np, Kutasov:1995ss}, which for Usp were proposed in \cite{Brodie:1996xm}. Namely one can turn on a superpotential term $tr(A^j)$ on the electric side, such term maps to a singlet on the magnetic, so a Higgsing process is induced. The study of this Higgsing might shed light on the dualities of \cite{Kutasov:1995ve, Kutasov:1995np, Kutasov:1995ss, Brodie:1996xm}. We expect the \emph{degenerate holomorphic operator ambiguity} encountered in this paper to play an important role.

There are quite a few \emph{self-dualities modulo flips} proposed in the literature \cite{Distler:1996ub, Csaki:1997cu, Karch:1997jp, Razamat:2018gbu}, such as $SU(2N)$ with antisymmetric, conjugate antisymmetric and $(4,4)$ fundamentals, or $SU(6)$ with $2$ antisymmetrics and $(2,6)$ fundamentals, or $SU(8)$ with $2$ antisymmetrics and $(0,8)$ fundamentals. A natural question is if such self-dualities can be proven using only the basic Seiberg and Intriligator-Pouliot dualities, as done in this work for $Usp(2N)$ with antisymmetric and $8$ fundamentals. Notice that many self-dual gauge theories have been constructed simply 'adding one flavor' to an S-confining gauge theory \cite{Csaki:1997cu, Karch:1997jp}, so the fact that the S-confining dualities can be proven \cite{Bajeot:2022kwt} is encouraging.

Related to the above point, many S-confinements and many self-dualities have been proposed for $4d$ $\cN=1$ $Spin(N)$ theories with spinors and vectors \cite{Csaki:1997cu, Karch:1997jp, Razamat:2018gbu, Sela:2019nqa}. It would be very interesting to find a way to deconfine spinorial matter and try to prove such proposals.

Limits to gauge theories with orthogonal and/or symplectic gauge groups. S-confinements for $3d$ $\cN=2$ theories with SO/Usp gauge groups and adjoint matter where recently proposed in \cite{Benvenuti:2021nwt}, and \cite{Amariti:2022wae} pointed out a relation to $4d$ $\cN=1$ S-confinements for Usp gauge group and antisymmetric matter. It might be interesting to deform the $4d$ $\cN=1$ sequential deconfinements as in  \cite{Amariti:2022wae}. See \cite{BenvenutiLoMonaco} for the sequential deconfinement of $3d$ $\cN=2$ rank-$2$ matter with SO/Usp gauge groups.

It would also be interesting to study \emph{degenerate holomorphic operator ambiguity} in other examples, possibly involving different kind of gauge groups.

\subsubsection*{Structure of the paper}
This paper is organized as follows.

\hspace{0.2cm}

In section \ref{BasicMoves} we recall our notation and the three basic dualities which will be iteratively used in the rest of the paper.

In section \ref{USp6} we discuss in detail a simple but non-trivial example, namely we sequentially deconfine $Usp(6)$ with an antisymmetric and $8$ fundamentals. This example is the simplest one where the issue of the \emph{degenerate holomorphic operator ambiguity} appears. We construct the fully deconfined dual, then we reconfine the quiver tail in order to prove the self-duality of the theory.

In section \ref{UspNF} we present the general sequential deconfinement of $Usp(2N)$ with an antisymmetric and $2F$ fundamentals.

In section \ref{reconf} we set $2F=8$, which allows to sequentially reconfine the quiver tail, and prove the self-duality modulo flips of the theory.

In section \ref{3d} we show how to reduce our $4d$ $\cN=1$ $Usp(2N)$ story to $3d$ $\cN=2$, re-obtaining the results of for $U(N)$ and $Usp(2N)$ found in \cite{Benvenuti:2020gvy}. Along the way we also derive new sequentially deconfined duals, namely for $U(N)$ with adjoint and $(F,F)$ fundamentals with monopole superpotentials.

\section{Basic S-confining and dualities moves} \label{BasicMoves}

In this section we present the basic ingredients that we use in this paper to obtain the more complicated dualities involving rank-2 matter field. The first basic move is the deconfinement of an antisymmetric field with a $Usp$ gauge group as in \cite{Bajeot:2022kwt, Bottini:2022vpy}, which is a modification of the original Berkooz deconfinement \cite{Berkooz:1995km}. This form was used in \cite{Bajeot:2022kwt, Bottini:2022vpy} to tackle Usp(2N) with $6$ fundamentals, which is S-confining.

Throughout the paper, blue $2N$ circles denote $Usp(2N)$ gauge nodes.

\noindent \textbf{$Usp$ deconfinement:}\footnote{On the r.h.s quiver in \eqref{UspDeconfinement}, we didn't include the flipper $\b_1$ on the drawing. Sometimes, as in \cite{Bajeot:2022kwt} this flipper is represented on the quiver by a cross, $\times$,  on the bifundamental field $b_1$. Moreover, the trace "$tr$" is taken using the $Usp$ invariant antisymmetric matrix. \\
Throughout the paper, a trace of an operator in the antisymmetric of a $Usp$ group is taken using the $Usp$ invariant antisymmetric matrix.}
\be \label{UspDeconfinement} \scalebox{0.9}{\bpic[node distance=2cm,gSUnode/.style={circle,red,draw,minimum size=8mm},gUSpnode/.style={circle,blue,draw,minimum size=8mm},fnode/.style={rectangle,draw,minimum size=8mm}]  
\node[gUSpnode] (G1) at (-3,0) {$2N$};
\node[fnode] (F1) at (-1,0) {$2F$};
\draw (G1) -- pic[pos=0.7,sloped]{arrow} (F1);
\draw (-2.7,0.4) to[out=90,in=0]  (-3,0.8) to[out=180,in=90] (-3.3,0.4);
\node[right] at (-2,-1.2) {$ \CW= 0$};
\node at (-2.5,0.9) {$A$};
\node at (0,0) {$\equiv$};
\node[gUSpnode] (G2) at (1.5,0) {$2N$};
\node[fnode] (F2) at (4,0) {$2F-1$};
\node[fnode,red] (F3) at (1.5,-1.8) {$1$};
\draw (G2) -- pic[pos=0.7,sloped]{arrow} (F2) node[midway,above] {$Q$};
\draw (G2) -- (F3) node[midway,left] {$P$};
\draw (1.8,0.4) to[out=90,in=0]  (1.5,0.8) to[out=180,in=90] (1.2,0.4);
\node at (2.1,0.9) {$A$};
\node at (5.5,0) {$\llra$};
\node[gUSpnode] (G3) at (7,0) {$2N$};
\node[gUSpnode] (G4) at (10.5,0) {$2N-2$};
\node[fnode] (F4) at (8.3,1.8) {$2F-1$};
\node[fnode] (F5) at (7,-2.5) {$1$};
\node[fnode,red] (F6) at (10.5,-2.5) {$1$};
\draw (G3) -- (G4);
\draw (G3) -- pic[pos=0.7,sloped]{arrow} (F4.south west);
\draw (G3) -- (F5) node[midway,left] {$v_1$};
\draw (G4) -- (F6) node[midway,right] {$v_1$};
\draw (F5) -- (F6) node[midway,below] {$h_1$};
\draw (F5.north east) -- (G4);
\node[right] at (5,-4) {$ \CW= v_1 b_1 d_1 + h_1 d_1 v_2 + \b_1 \, \tr(b_1 b_1) $};
\node at (8.7,0.3) {$b_1$};
\node at (8.5,-1) {$d_1$};
\node at (6.9,1.1) {$Q$};
\epic} \ee

In order to get this deconfinement we have to use another basic move. It is the S-confining result of Intriligator-Pouliot (IP) \cite{Intriligator:1995ne} that involves $Usp(2N)$ gauge group and $2N+4$ fields in the fundamental representation. 

\noindent \textbf{IP S-confining duality:}
\be \label{IPconfining} \bpic[node distance=2cm,gSUnode/.style={circle,red,draw,minimum size=8mm},gUSpnode/.style={circle,blue,draw,minimum size=8mm},fnode/.style={rectangle,draw,minimum size=8mm}]  
\node[gUSpnode] (G1) at (-3,0) {$2N$};
\node[fnode] (F1) at (-0.5,0) {$2N+4$};
\draw (G1) -- pic[pos=0.7,sloped]{arrow} (F1);
\node[right] at (-2.7,-1.2) {$ \CW= 0$};
\node at (-1.9,0.4) {$Q$};
\node at (1.5,0) {$\longleftrightarrow$};
\node[fnode] (F2) at (3.5,0) {$2N+4$};
\draw (4,0.4) to[out=90,in=0] pic[pos=0.1,sloped]{arrow} (3.5,1) to[out=180,in=90] pic[pos=0.7,sloped,very thick]{arrow=latex reversed} (3,0.4);
\node[right] at (2.3,-1.2) {$ \CW= \Pf( \mu)$};
\node at (4.2,1) {$\mu$};
\node at (-0.5,-2) {Mapping: $\qquad\qquad tr(Q Q) \longleftrightarrow \mu$};
\epic \ee 
The last move concerns the same $Usp$ theory but with $2F \ge 2N+4$ fundamentals now. It is the IP duality \cite{Intriligator:1995ne}. In quiver notation it reads

\noindent \textbf{IP duality:}
\be \label{IPduality} \bpic[node distance=2cm,gSUnode/.style={circle,red,draw,minimum size=8mm},gUSpnode/.style={circle,blue,draw,minimum size=8mm},fnode/.style={rectangle,draw,minimum size=8mm}]  
\node[gUSpnode] (G1) at (-3,0) {$2N$};
\node[fnode,minimum size = 0.9cm] (F1) at (-1,0) {$2F$};
\draw (G1) -- pic[pos=0.7,sloped]{arrow} (F1);
\node[right] at (-2.8,-1.5) {$ \CW= 0$};
\node at (-2,0.4) {$Q$};
\node at (0.5,0) {$\llra$};
\node[gUSpnode] (G2) at (3,0) {\scalebox{0.7}{$2F-2N-4$}};
\node[fnode,minimum size = 0.9cm] (F2) at (5.7,0) {$2F$};
\draw (G2) -- pic[pos=0.7,sloped,very thick]{arrow=latex reversed} (F2);
\draw (6,0.5) to[out=90,in=0] pic[pos=0.1,sloped]{arrow} (5.7,1) to[out=180,in=90] pic[pos=0.6,sloped,very thick]{arrow=latex reversed} (5.4,0.5);
\node[right] at (3.2,-1.5) {$ \CW= tr(q \, q) \, \P $};
\node at (4.6,0.3) {$q$};
\node at (6,1.2) {$\P$};
\node at (-1.7,-2.5) {Mapping: $\qquad\qquad tr(Q Q) \longleftrightarrow \P$};
\epic \ee

\section{Case study: $Usp(6)$ with ${\tiny\ydiagram{1,1}} $ + $8 \, {\tiny\ydiagram{1}}$} \label{USp6}
In this section we  study the $Usp(6)$ gauge theory with matter in the antisymmetric representation and $8$ fundamentals, that is $4$ flavors. It is the simplest example that  contains all the ingredients that we want to exhibit. In the next sections, we will show the general case. It is known that this theory is self-dual \emph{modulo flips} \cite{Csaki:1996eu}. We will prove the self-duality using the basic moves of the previous section \eqref{UspDeconfinement}, \eqref{IPconfining} and \eqref{IPduality}.

\subsection{Sequential deconfinement}
The first step is the deconfinement of the antisymmetric with \eqref{UspDeconfinement}. We get the following two frames $\CT_{0}$ and $\CT_{0'}$
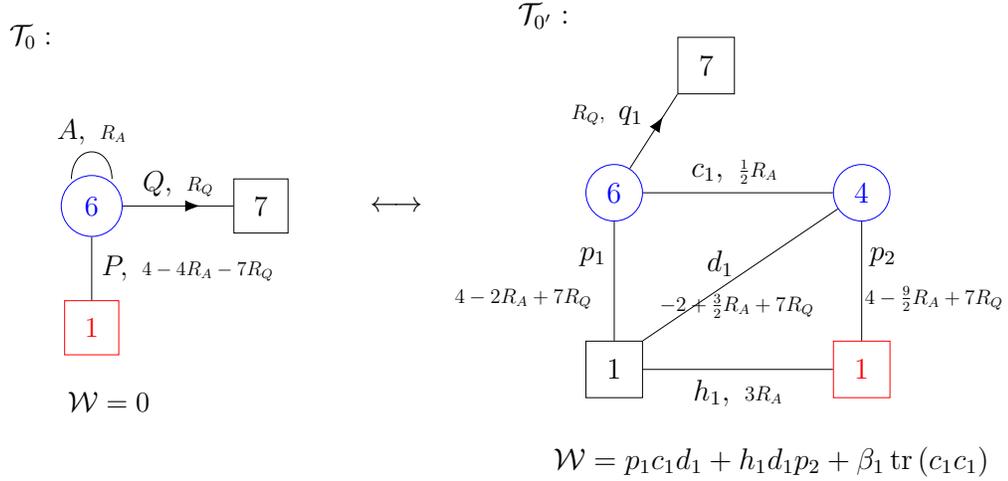
\begin{center}
\scalebox{0.9}{\bpic[node distance=2cm,gSUnode/.style={circle,red,draw,minimum size=8mm},gUSpnode/.style={circle,blue,draw,minimum size=8mm},fnode/.style={rectangle,draw,minimum size=8mm}]  
\node at (1.6,2.5) {$\CT_{0}:$};
\node[gUSpnode,minimum size=9mm] (G2) at (2.5,0) {$6$};
\node[fnode] (F2) at (5,0) {$7$};
\node[fnode,red] (F3) at (2.5,-1.8) {$1$};
\draw (G2) -- pic[pos=0.7,sloped]{arrow} (F2) node[midway,above] {$Q, \,$ \scalebox{0.7}{$R_Q$}};
\draw (G2) -- (F3) node[midway,right] {$P, \,$ \scalebox{0.7}{$4 -4 R_A - 7 R_Q$}};
\draw (2.8,0.4) to[out=90,in=0]  (2.5,0.8) to[out=180,in=90] (2.2,0.4);
\node[right] at (2,-2.9) {$ \CW= 0$};
\node at (2.5,1.1) {$A, \,$ \scalebox{0.7}{$R_A$}};
\node at (7,0) {$\llra$};
\node at (0,-4) {};
\epic} 
\hspace{-0.1cm}
\scalebox{0.94}{\bpic[node distance=2cm,gSUnode/.style={circle,red,draw,minimum size=8mm},gUSpnode/.style={circle,blue,draw,minimum size=8mm},fnode/.style={rectangle,draw,minimum size=8mm}]  
\node at (6,2.5) {$\CT_{0'}:$};
\node[gUSpnode] (G3) at (7,0) {$6$};
\node[gUSpnode] (G4) at (10.5,0) {$4$};
\node[fnode] (F4) at (8.3,1.8) {$7$};
\node[fnode] (F5) at (7,-2.5) {$1$};
\node[fnode,red] (F6) at (10.5,-2.5) {$1$};
\draw (G3) -- (G4);
\draw (G3) -- pic[pos=0.7,sloped]{arrow} (F4.south west);
\draw (G3) -- (F5);
\draw (G4) -- (F6);
\draw (F5) -- (F6) node[midway,below] {$h_1, \,$ \scalebox{0.7}{$3 R_A$}};
\draw (F5.north east) -- (G4);
\node at (6.7,-0.9) {$p_1$};
\node[right] at (4.6,-1.5) {\scalebox{0.7}{$4 - 2R_A + 7R_Q$}};
\node at (10.8,-0.9) {$p_2$};
\node[right] at (10.4,-1.5) {\scalebox{0.7}{$4 - \frac{9}{2} R_A + 7R_Q$}};
\node[right] at (6,-3.8) {$ \CW= p_1 c_1 d_1 + h_1 d_1 p_2 + \b_1 \, \tr(c_1 c_1) $};
\node at (8.7,0.3) {$c_1, \,$ \scalebox{0.7}{$\frac{1}{2} R_A$}};
\node at (8.5,-1) {$d_1$};
\node[right] at (7.5,-1.6) {\scalebox{0.7}{$-2 + \frac{3}{2} R_A + 7 R_Q$}};
\node at (6.9,1.1) {\scalebox{0.7}{$R_Q, \,$} $q_1$};
\epic} 
\end{center}
The global symmetry of the original theory is $SU(8) \times U(1)_A \times U(1)_R$. When we split the $8$ chirals into $7+1$, we split the $SU(8)$ into $SU(7) \times U(1)_P$. So in the splitted form, the R-charges of the fields should be a function of two variables (for the two global $U(1)$'s that can mix with the $U(1)_R$). We choose to write the R-charges in terms of $R_Q$ and $R_A$. Then the R-charges of the other fields are determined by the $U(1)_R$ ABJ anomaly\footnote{This is the same as requiring the vanishing of the NSVZ $\b$ function.} and by the requirement that any superpotential term should have R-charge equal to 2. We have written the R-charges of the fields next to them. 

Then we dualize the $Usp(6)$ node with \eqref{IPduality}. The fields $d_1$ and $\b_1$ get a mass. After integrating them out and rescaling the fields to put a $+1$ in front of each term in the superpotential, we get

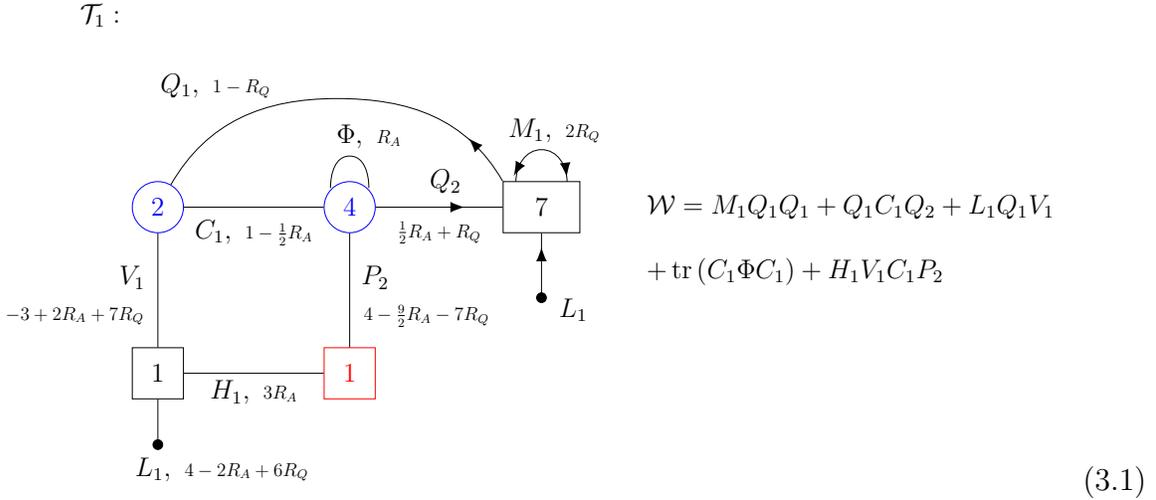
\be \label{Usp6T1} \scalebox{0.85}{\bpic[node distance=2cm,gSUnode/.style={circle,red,draw,minimum size=8mm},gUSpnode/.style={circle,blue,draw,minimum size=8mm},fnode/.style={rectangle,draw,minimum size=8mm}]  
\node at (-2.4,3) {$\CT_{1}:$};
\node[gUSpnode] (G1) at (-1.5,0) {$2$};
\node[gUSpnode] (G2) at (1.5,0)  {$4$};
\node[fnode, minimum width=1.2cm] (F1) at (4.5,0) {$7$};
\node[fnode] (F2) at (-1.5,-2.6) {$1$};
\node[fnode,red] (F3) at (1.5,-2.6) {$1$};
\draw (G1) -- (G2);
\draw (G1) to[out=60,in=180] (1.3,1.7) to[out=0,in=120] pic[pos=0.8,sloped,very thick]{arrow=latex reversed} (F1.north west);
\draw (G1) -- (F2);
\draw (G2) -- (F3);
\draw (G2) -- pic[pos=0.7,sloped]{arrow} (F1); 
\draw (F2) -- (F3);
\draw[-*] (F2) -- (-1.5,-3.8);
\draw[*-] (4.5,-1.5) -- pic[pos=0.8,sloped]{arrow} (F1);
\draw (1.8,0.3) to[out=90,in=0]  (1.5,0.8) to[out=180,in=90] (1.2,0.3);
\draw (4.9,0.4) to[out=90,in=0] pic[pos=0.1,sloped]{arrow} (4.5,0.9) to[out=180,in=90] pic[pos=0.6,sloped,very thick]{arrow=latex reversed} (4.1,0.4);
\node[right] at (6,0) {$\CW= M_1 Q_1 Q_1 + Q_1 C_1 Q_2 + L_1 Q_1 V_1$};
\node[right] at (6,-1) {$+ \, \tr(C_1 \P C_1) + H_1 V_1 C_1 P_2$};
\node at (-0.5,-4.1) {$L_1, \,$ \scalebox{0.7}{$4 - 2R_A + 6R_Q$}};
\node at (5,-1.6) {$L_1$};
\node at (-0.6,1.9) {$Q_1, \,$ \scalebox{0.7}{$1-R_Q$}};
\node at (3,0.4) {$Q_2$};
\node at (2.9,-0.4) {\scalebox{0.7}{$\frac{1}{2} R_A + R_Q$}};
\node at (-1.9,-1.1) {$V_1$};
\node at (-2.8,-1.7) {\scalebox{0.7}{$-3 + 2 R_A + 7R_Q$}};
\node at (1.9,-1.1) {$P_2$};
\node at (2.7,-1.7) {\scalebox{0.7}{$4 - \frac{9}{2} R_A - 7R_Q$}};
\node at (0,-0.4) {$C_1, \,$ \scalebox{0.7}{$1 - \frac{1}{2}R_A$}};
\node at (0,-2.9) {$H_1, \,$ \scalebox{0.7}{$3 R_A$}};
\node at (4.7,1.2) {$M_1, \,$ \scalebox{0.7}{$2R_Q$}};
\node at (1.8,1.1) {$\P, \,$ \scalebox{0.7}{$R_A$}};
\epic} \ee
In $\CT_1$ the antisymmetric field, $\P$, is traceless because the trace component has been killed by the equation of motion (E.O.M) of the flipper $\b_1$. The mapping of the chiral ring generators after this first step is the following
\be \label{map1Usp6}
\scalebox{0.9}{$
\ba{c}\CT_0 \\
Q Q \\
\tr(Q A^i Q) \\
\tr(Q A^j P) \\
\tr(Q A^{2} P) \\
\tr(A^2) \\
\tr(A^{3})
\ea
\Longleftrightarrow
\ba{c} \CT_{0'} \\
q_{1} q_{1} \\
\tr(q_{1} \, (c_1 c_1)^i  \, q_{1}) \\
\tr(q_{1} \, c_1 (c_1 c_1)^j \, p_{2}) \\
q_{1} \, p_1 \\
\tr((c_1 c_1)^2) \\
h_1
\ea
\Longleftrightarrow
\ba{c}\CT_1 \\
M_{1} \\
\tr(Q_{2} \,\P^{i-1} \, Q_{2}) \\
\tr(Q_{2} \, \P^{j} \, P_{2}) \\
L_1 \\
\tr(\P^2) \\
H_1
\ea
\qquad
\ba{l}\\
\\
i=1, 2 \\
j=0, 1 \\
\\
\\
\\
\ea
$}
\ee
It can be checked that the mapping \eqref{map1Usp6} is consistent with the R-charges of the operators. Now we iterate the procedure. 

We now deconfine the traceless antisymmetric field $\P$.

\be \label{Usp6T1'} \scalebox{0.83}{\bpic[node distance=2cm,gSUnode/.style={circle,red,draw,minimum size=8mm},gUSpnode/.style={circle,blue,draw,minimum size=8mm},fnode/.style={rectangle,draw,minimum size=8mm}]  
\node at (-2.5,3.5) {$\CT_{1'}:$};
\node[gUSpnode] (G1) at (-1.5,0) {$2$};
\node[gUSpnode] (G2) at (1.5,0) {$4$};
\node[gUSpnode] (G3) at (4.5,0) {$2$};
\node[fnode,minimum width=1.2cm] (F1) at (1.5,2.3) {$7$};
\node[fnode] (F2) at (-1.5,-2.6) {$1$};
\node[fnode] (F3) at (1.5,-2.6) {$1$};
\node[fnode,red] (F4) at (4.5,-2.6) {$1$};
\draw (G1) -- (G2);
\draw (G2) -- (G3);
\draw (G1.north east) -- pic[pos=0.8,sloped,very thick]{arrow=latex reversed} (F1.south west);
\draw (G1) -- (F2);
\draw (G2) -- pic[pos=0.8,sloped]{arrow} (F1);
\draw (G2) -- (F3);
\draw (G3) -- (F4);
\draw (G3) -- (F3.north east);
\draw (F3) -- (F4);
\draw[-*] (F2) -- (-1.5,-3.8);
\draw[*-] (3.2,2.3) -- pic[pos=0.6,sloped,very thick]{arrow=latex reversed} (F1);
\draw (1.9,2.7) to[out=90,in=0] pic[pos=0.1,sloped]{arrow} (1.5,3.2) to[out=180,in=90] pic[pos=0.6,sloped,very thick]{arrow=latex reversed} (1.1,2.7);
\draw (F2.south east) to[out=-60,in=180] (1.5,-3.8) to[out=0,in=-120] (F4.south west);
\node[right] at (7.2,0) {$\CW= m_1 q_1 q_1 + q_1 c_1 q_2 + l_1 v_1 q_1$};
\node[right] at (7.2,-1) {$+ \, \tr(c_1 c_2 c_2 c_1) + h_1 v_1 c_1 c_2 p_3$};
\node[right] at (7.2,-2) {$+ \, h_2 d_2 p_3 + p_2 c_2 d_2 + \b_2 \, \tr(c_2 c_2)$};
\node at (-0.4,-4.2) {$l_1, \,$ \scalebox{0.7}{$4- 2 R_A -6R_Q$}};
\node at (3.6,2.2) {$l_1$};
\node at (2.8,-1.1) {$d_2$};
\node at (3.1,-1.6) {\scalebox{0.7}{$-2+3 R_A +7R_Q$}};
\node at (-0.8,1.4) {\scalebox{0.7}{$1-R_Q, \,$} $q_1$};
\node at (2.7,1.2) {$q_2, \, \,$\scalebox{0.7}{$\frac{1}{2} R_A + R_Q$}};
\node at (-1.8,-0.9) {$v_1$};
\node at (-2.8,-1.6) {\scalebox{0.7}{$-3 + 2 R_A + 7 R_Q$}};
\node at (1.2,-0.9) {$p_2$};
\node at (0.3,-1.6) {\scalebox{0.7}{$-4 - \frac{7}{2} R_A - 7 R_Q$}};
\node at (4.8,-0.9) {$p_3$};
\node at (5.7,-1.6) {\scalebox{0.7}{$-4 - 5 R_A - 7 R_Q$}};
\node at (0,0.3) {$c_1$};
\node at (0,-0.3) {\scalebox{0.7}{$1-\frac{1}{2} R_A$}};
\node at (3,0.3) {$c_2$};
\node at (3,-0.4) {\scalebox{0.7}{$\frac{1}{2} R_A$}};
\node at (0.2,-3.4) {$h_1, \,$ \scalebox{0.7}{$3 R_A$}};
\node at (3,-2.9) {$h_2, \,$ \scalebox{0.7}{$2 R_A$}};
\node at (2.1,3.4) {$m_1, \,$ \scalebox{0.7}{$2 R_Q$}};
\epic} \ee
Now we dualize the $Usp(4)$ node. There will not be any antisymmetric field for the other gauge groups because they are $Usp(2)$\footnote{The dualization creates however singlets that correspond to the trace part of the would be antisymmetric. We call the singlet on the left $\a_1$. The one on the right will receive a mass with the singlet $\b_2$ so we do not need to give it a name.}. The fields $q_1$, $d_2$ and $\b_2$ get a mass. Moreover, $\tr(c_1 c_2 c_2 c_1)$ becomes a mass term therefore there will be no link between the two $Usp(2)$ gauge groups. After integrating these massive fields  out and a rescaling we get
\be \label{Usp6T2} \scalebox{0.84}{\bpic[node distance=2cm,gSUnode/.style={circle,red,draw,minimum size=8mm},gUSpnode/.style={circle,blue,draw,minimum size=8mm},fnode/.style={rectangle,draw,minimum size=8mm}]  
\node at (-1,2.3) {$\CT_{2}:$};
\node[gUSpnode] (G1) at (0,0) {$2$};
\node[gUSpnode] (G2) at (3,0) {$4$};
\node[gUSpnode] (G3) at (6,0) {$2$};
\node[fnode,minimum width=1.4cm] (F1) at (9.5,0) {$7$};
\node[fnode] (F2) at (0,-2.8) {$1$};
\node[fnode] (F3) at (3,-2.8) {$1$};
\node[fnode,red] (F4) at (6,-2.8) {$1$};
\draw (G1) -- (G2);
\draw (G2) -- (G3);
\draw (G2) to[out=60,in=180] (5.8,1.6) to[out=0,in=120] pic[pos=0.8,sloped,very thick]{arrow=latex reversed} (F1.north west);
\draw (G1) -- (F2);
\draw (G2) -- (F3);
\draw (G3) -- pic[pos=0.8,sloped]{arrow} (F1);
\draw (G3) -- (F4);
\draw (G1) -- (F3.north west);
\draw[-*] (F2) -- (0,-4);
\draw[*-] (9.2,-1.4) -- pic[pos=0.8,sloped]{arrow} (9.2,-0.4);
\draw[-*] (F3) -- (3,-4);
\draw[*-] (9.9,-1.4) -- pic[pos=0.8,sloped]{arrow} (9.9,-0.4);
\draw (9.9,0.4) to[out=90,in=0] pic[pos=0.1,sloped]{arrow} (9.5,0.9) to[out=180,in=90] pic[pos=0.6,sloped,very thick]{arrow=latex reversed} (9.1,0.4);
\draw (10.1,0.4) to[out=90,in=0] pic[pos=0.2,sloped]{arrow} (9.5,1.2) to[out=180,in=90] pic[pos=0.6,sloped,very thick]{arrow=latex reversed} (8.9,0.4);
\draw (F2.south east) to[out=-60,in=180] (3,-4.6) to[out=0,in=-120] (F4.south west);
\node[right] at (-3.5,-5.5) {$\CW= M_1 \tr(B_1 Q_2 Q_2 B_1) + M_2 Q_2 Q_2 + \tr(C_2 \P  C_2) + \a_1 \, \tr(B_1 B_1) + \tr(B_1 C_2 C_2 B_1)$};
\node[right] at (-3.5,-6.5) {$+ \, R_1 B_1 V_2 + L_1 V_1 B_1 Q_2 + L_2 V_2 Q_2 + Q_2 C_2 Q_3 + H_1 V_1 B_1 C_2 P_3 + H_2 ?$};
\node at (9.1,-1.7) {$L_1$};
\node at (0.4,-4) {$L_1$};
\node at (10.1,-1.7) {$L_2$};
\node at (3.4,-4) {$L_2$};
\node at (1.7,-1.2) {$R_1$};
\node at (1.7,-1.9) {\scalebox{0.7}{$5-4R_A-7R_Q$}};
\node at (-0.4,-1.2) { $V_1$};
\node at (-1.2,-1.9) {\scalebox{0.7}{$-3+2R_A+7R_Q$}};
\node at (3.3,-1.2) {$V_2$};
\node at (4.1,-1.9) {\scalebox{0.7}{$-3+\frac{7}{2}R_A+7R_Q$}};
\node at (6.3,-1.2) {$P_3$};
\node at (7.1,-1.9) {\scalebox{0.7}{$4-5R_A-7R_Q$}};
\node at (1.5,0.3) {$B_1, \,$ \scalebox{0.7}{$\frac{1}{2} R_A$}};
\node at (4.5,-0.3) {$C_2, \,$ \scalebox{0.7}{$1-\frac{1}{2} R_A$}};
\node at (1.7,-3.8) {$H_1, \,$ \scalebox{0.7}{$3 R_A$}};
\node at (7,-4) {\textcolor{Orange}{+ singlet $H_2, \,$ \scalebox{0.7}{$2 R_A$}}};
\node at (9.8,1.5) {$M_1,M_2$};
\node at (5.6,1.1) {$Q_2, \,$ \scalebox{0.7}{$1-\frac{1}{2} R_A-R_Q$}};
\node at (7.5,-0.4) {$Q_3, \,$ \scalebox{0.7}{$R_A+R_Q$}};
\epic} \ee
At this step, we face a feature that we call the \emph{degenerate holomorphic operator ambiguity}. It arises when we ask what is the operator flipped by the singlet $H_2$.

\subsection{Degenerate holomorphic operator ambiguity}
If we apply the rules of Seiberg duality locally in the quiver, as is usually done, using the mapping \eqref{IPduality}, we would conclude that it is $\textcolor{Violet}{\cO_1 = V_2 C_2 P_3}$, so the superpotential should contain $H_2 \textcolor{Violet}{\cO_1}$. However, for the quiver at hand, there is another candidate, $\textcolor{ForestGreen}{\cO_2 = V_1 R_1}$. Both $\cO_1$ and $\cO_2$ are gauge singlets, singlets under the $SU(7)$ flavor symmetry and have R-charge: $R(\textcolor{ForestGreen}{V_1 R_1}) = R(\textcolor{Violet}{V_2 C_2 P_3}) = 2-2 R(A)$ (which implies that the two operators have the same charges under the flavor $U(1)'s$). Therefore they are degenerate operators.\footnote{Let us emphasize that there is no ambiguity with $H_1$ because there is only one operator which is singlet under the gauge symmetry, the $SU(7)$ flavor symmetry and has R-charge equal to $2-3 R(A)$. This operator is $V_1 B_1 C_2 P_3$.}. So it might be that the precise operator flipped by $H_2$ is not exactly $\cO_1$, but some linear combination of $\cO_1$ and $\cO_2$. 

We claim that the correct answer is that the operator flipped by $H_2$ is $\textcolor{ForestGreen}{\cO_2 = V_1 R_1}$, instead of the naive $\cO_1$. Our argument in favor of this statement comes from dualizing some nodes in the quiver, as we will explain soon. So in a sense the fact the the correct operator is not the naive one is due to quantum relations, which become classical relation after Seiberg duality.


Our strategy to decide the correct operator is to use dualities in order to go to a frame where F-term equations can answer the question. In this case, we apply IP S-confining duality \eqref{IPconfining} on the left $Usp(2)$ gauge node of theory $\cT_2$ \emph{with the singlet $H_2$ removed}. We consider the same theory with the flipping removed, the question becomes which linear combination of the two degenerate holomorphic operators $\cO_1$ and $\cO_2$ is non zero in the chiral ring. Since the answer, as we will see, is that $\cO_2$ is non zero in the chiral ring, the quantum relation in the unflipped theory is $\cO_1=0$. 

The $Usp(2)$ we dualize is coupled to $6$ fundamentals and so it confines, producing a traceless antisymmetric field $B$ (the trace part is killed by the flipper $\a_1$).  We get
\be \label{Switching2} \scalebox{0.85}{\bpic[node distance=2cm,gSUnode/.style={circle,red,draw,minimum size=8mm},gUSpnode/.style={circle,blue,draw,minimum size=8mm},fnode/.style={rectangle,draw,minimum size=8mm}]  
\node[gUSpnode] (G2) at (3,0) {$4$};
\node[gUSpnode] (G3) at (6,0) {$2$};
\node[fnode,minimum width=1.4cm] (F1) at (9,0) {$7$};
\node[fnode] (F2) at (0,-2.4) {$1$};
\node[fnode] (F3) at (3,-2.4) {$1$};
\node[fnode,red] (F4) at (6,-2.4) {$1$};
\node[right] at (-3.5,-5.1) {$\CW= M_1 \tr(Q_2 B Q_2) + M_2 Q_2 Q_2 + \tr(C_2 B  C_2) + X V_2 + L_1 P Q_2 + L_2 V_2 Q_2 + Q_2 C_2 Q_3$};
\node[right] at (-3.5,-7.1) {$+ \, H_1 P C_2 P_3 + H_2 ? + \Pf$ \scalebox{0.7}{$\begin{pmatrix}
B & \vdots & P & \vdots & X \\
\dots & \dots & \dots & \dots & \dots \\
& \vdots & 0 & \vdots & s \\
& & \dots & \dots & \dots\\
& & & \vdots & 0
\end{pmatrix}$}};
\draw (G2) -- (G3);
\draw (3.4,0.2) to[out=60,in=180] (5.8,1.6) to[out=0,in=120] pic[pos=0.8,sloped,very thick]{arrow=latex reversed} (F1.north west);
\draw (G2) -- (F2.north east);
\draw (F2) -- (F3);
\draw (G3) -- pic[pos=0.8,sloped]{arrow} (F1);
\draw (G3) -- (F4);
\draw (3.2,-0.3) -- (3.2,-2);
\draw (2.8,-0.3) -- (2.8,-2);
\draw[-*] (F2) -- (0,-3.6);
\draw[*-] (8.7,-1.4) -- pic[pos=0.8,sloped]{arrow} (8.7,-0.4);
\draw[-*] (F3) -- (3,-3.6);
\draw[*-] (9.4,-1.4) -- pic[pos=0.8,sloped]{arrow} (9.4,-0.4);
\draw (3.3,0.3) to[out=90,in=0] (3,0.8) to[out=180,in=90] (2.7,0.3);
\draw (9.4,0.4) to[out=90,in=0] pic[pos=0.1,sloped]{arrow} (9,0.9) to[out=180,in=90] pic[pos=0.6,sloped,very thick]{arrow=latex reversed} (8.6,0.4);
\draw (9.6,0.4) to[out=90,in=0] pic[pos=0.2,sloped]{arrow} (9,1.2) to[out=180,in=90] pic[pos=0.6,sloped,very thick]{arrow=latex reversed} (8.4,0.4);
\draw (F2.south east) to[out=-60,in=180] (3,-4.2) to[out=0,in=-120] (F4.south west);
\node at (2.5,0.9) {$B$};
\node at (8.6,-1.7) {$L_1$};
\node at (0.4,-3.6) {$L_1$};
\node at (9.6,-1.7) {$L_2$};
\node at (3.4,-3.6) {$L_2$};
\node at (1.5,-0.7) {$P$};
\node at (0,-1.2) {\scalebox{0.7}{$-3+\frac{5}{2}R_A+7R_Q$}};
\node at (2.5,-1.4) {$X$};
\node at (3.6,-1.4) {$V_2$};
\node at (6.3,-1.3) {$P_3$};
\node at (4.7,-0.3) {$C_2$};
\node at (1.6,-3.6) {$H_1$};
\node at (6.8,-3.6) {\textcolor{Orange}{+ singlet $H_2$}};
\node at (9.8,1.5) {$M_1,M_2$};
\node at (4.4,0.8) {$Q_2$};
\node at (7.5,-0.4) {$Q_3$};
\node at (1.6,-2.2) {$s, \,$ \scalebox{0.7}{$2-2R_A$}};
\epic} \ee

\noindent The Pfaffian term gives: $\Pf$ \scalebox{0.7}{$\begin{pmatrix}
B & \vdots & P & \vdots & X \\
\dots & \dots & \dots & \dots & \dots \\
& \vdots & 0 & \vdots & s \\
& & \dots & \dots & \dots\\
& & & \vdots & 0
\end{pmatrix}$} $\sim \e_4 \, (B^2) s + \e_4 \, (B P X)$

\noindent The fields $X$ and $V_2$ are massive. The E.O.M of $X$ gives: $V_2 + B P = 0$.

\noindent In addition, the F-term equation for the singlet $H_1$ gives: $P C_2 P_3 = 0$.

Combining these two informations, we can resolve the ambiguity about the operators $\textcolor{Violet}{\cO_1}$ and $\textcolor{ForestGreen}{\cO_2}$. Indeed in this frame these operators become
\begin{align}
\textcolor{Violet}{\cO_1 = V_2 C_2 P_3 \quad } &\textcolor{Violet}{\longrightarrow \quad V_2 C_2 P_3 \, \,} \overset{\textcolor{red}{\text{{\scriptsize E.O.M $X$}}}}{\textcolor{Violet}{=}} \textcolor{Violet}{\, \, B P C_2 P_3 \, \,} \overset{\textcolor{red}{\text{{\scriptsize F-term $H_1$}}}}{\textcolor{Violet}{\simeq}} \textcolor{Violet}{\, \, 0} \label{SwitchingOp} \\
\textcolor{ForestGreen}{\cO_2 = V_1 R_1 \quad} &\textcolor{ForestGreen}{\longrightarrow \quad s}
\end{align}
The symbol $\simeq$ in the last of equality in \eqref{SwitchingOp} means an equivalence in the chiral ring.

Therefore we conclude that the non-zero operator in the chiral ring is $\textcolor{ForestGreen}{\cO_2 = V_1 R_1}$ and so it should be this one that enters in the superpotential with $H_2$.

\subsection{Fully deconfined frame}
We now go back to our deconfining procedure, and dualize the right $Usp(2)$ node in $\CT_2$ using \eqref{IPduality}. We reach a frame that we call  \say{fully deconfined} as in \cite{Benvenuti:2020gvy}.
\be \label{Usp6TDec} \scalebox{0.85}{\bpic[node distance=2cm,gSUnode/.style={circle,red,draw,minimum size=8mm},gUSpnode/.style={circle,blue,draw,minimum size=8mm},fnode/.style={rectangle,draw,minimum size=8mm}]  
\node at (-1,2.3) {$\CT_{Dec}:$};
\node[gUSpnode] (G1) at (0,0) {$2$};
\node[gUSpnode] (G2) at (3,0) {$4$};
\node[gUSpnode] (G3) at (6,0) {$6$};
\node[fnode,minimum width=1.4cm] (F1) at (9.5,0) {$7$};
\node[fnode] (F2) at (0,-2.8) {$1$};
\node[fnode] (F3) at (3,-2.8) {$1$};
\node[fnode,red] (F4) at (6,-2.8) {$1$};
\draw (G1) -- (G2);
\draw (G2) -- (G3);
\draw (G1) -- (F2);
\draw (G2) -- (F3);
\draw (G3) -- pic[pos=0.8,sloped,very thick]{arrow=latex reversed} (F1);
\draw (G2) -- (F4);
\draw (G3) -- (F4);
\draw (G1) -- (F3.north west);
\draw (F2) -- (F3);
\draw[-*] (F2) -- (0,-4);
\draw[*-] (9.1,-1.4) -- pic[pos=0.8,sloped]{arrow} (9.1,-0.4);
\draw[-*] (F3) -- (3,-4);
\draw[*-] (10,-1.4) -- pic[pos=0.8,sloped]{arrow} (10,-0.4);
\draw[-*] (F4) -- (6,-4);
\draw (9.9,0.4) to[out=90,in=0] pic[pos=0.1,sloped]{arrow} (9.5,0.9) to[out=180,in=90] pic[pos=0.6,sloped,very thick]{arrow=latex reversed} (9.1,0.4);
\draw (9.5,1.3) node {$\vdots$};
\draw (10.1,0.4) to[out=90,in=0] pic[pos=0.2,sloped]{arrow} (9.5,1.7) to[out=180,in=90] pic[pos=0.6,sloped,very thick]{arrow=latex reversed} (8.9,0.4);
\draw (3.3,0.3) to[out=90,in=0] (3,0.8) to[out=180,in=90] (2.7,0.3);
\draw (F2.south east) to[out=-60,in=180] (3,-4.6) to[out=0,in=-120] (F4.south west);
\node[right] at (-3.5,-5.5) {$\CW= m_1 \tr(b_1 b_2 q_3 q_3 b_2 b_1) + m_2 \tr(b_2 q_3 q_3 b_2) + m_3 q_3 q_3 + \a_1 \, \tr(b_1 b_1) + \a_2 \, \tr(b_2 b_2) $};
\node[right] at (-3.5,-6.5) {$+ \, \tr(b_1 a_2  b_1) + \tr(b_2 a_2 b_2) + r_1 b_1 v_2 + r_2 b_2 v_3 + l_1 v_1 b_1 b_2 q_3 + l_2 v_2 b_2 q_3 + l_3 v_3 q_3 + h_1 v_1 b_1 r_2+ h_2 v_1 r_1$};
\node at (9,-1.7) {$l_1$};
\node at (0.4,-4) {$l_1$};
\node at (9.6,-0.8) {$\dots$};
\node at (10.2,-1.7) {$l_3$};
\node at (3.4,-4) {$l_2$};
\node at (6.4,-4) {$l_3$};
\node at (1.7,-1.2) {$r_1$};
\node at (4.7,-1.2) {$r_2$};
\node at (3.5,0.9) {$a_2$};
\node at (1.7,-1.9) {\scalebox{0.7}{$5-4R_A-7R_Q$}};
\node at (-0.4,-1.2) { $v_1$};
\node at (-1.2,-1.9) {\scalebox{0.7}{$-3+2R_A+7R_Q$}};
\node at (3.3,-1.2) {$v_2$};
\node at (4.1,-1.9) {\scalebox{0.7}{$-3+\frac{7}{2}R_A+7R_Q$}};
\node at (6.3,-1.2) {$v_3$};
\node at (7.1,-1.9) {\scalebox{0.7}{$4-5R_A-7R_Q$}};
\node at (1.5,0.3) {$b_1, \,$ \scalebox{0.7}{$\frac{1}{2} R_A$}};
\node at (4.5,0.3) {$b_2, \,$ \scalebox{0.7}{$\frac{1}{2} R_A$}};
\node at (1.7,-3.8) {$h_1, \,$ \scalebox{0.7}{$3 R_A$}};
\node at (1.7,-3) {$h_2, \,$ \scalebox{0.7}{$2 R_A$}};
\node at (9.8,2) {$m_1, \dots, m_3$};
\node at (7.5,-0.4) {$q_3, \,$ \scalebox{0.7}{$R_A+R_Q$}};
\epic} \ee
The antisymmetric field $a_2$ is traceless, as all the antisymmetric field of $Usp$ that will appear in this paper. Once again there is the question of the operator flipped by $h_2$ because $v_2 \, r_2$ has the same quantum numbers as $v_1 \, r_1$. Using the same procedure of confining from the left, we would obtain that the operator $v_2 r_2$ is $0$ on the chiral ring. Therefore we claim that the correct final superpotential is the one with this switching procedure and not the one that we would have got using naive iteration of IP dualities. The final mapping of the chiral ring generators is
\be \label{map2Usp6}
\scalebox{0.9}{$
\ba{c}\CT_1 \\
M_{1} \\
Q_{2} \, Q_{2} \\
\tr(Q_{2} \, \P \, Q_{2}) \\
Q_{2} \, P_{2} \\
\tr(Q_{2} \, \P \, P_{2}) \\
L_1 \\
\tr(\P^{2}) \\
H_1
\ea
\Longleftrightarrow
\ba{c} \CT_{1'} \\
m_{1} \\
q_{2} \, q_{2} \\
\tr(q_{2} \, (c_2 c_2) \, q_{2}) \\
q_{2} \, c_2 \, p_{3} \\
q_{2} \, p_2 \\
l_1 \\
h_2 \\
h_1 
\ea
\Longleftrightarrow
\ba{c} \CT_{2} \\
M_{1} \\
M_{2} \\
Q_{3} \, Q_{3} \\
Q_{3} \, P_3 \\
L_2 \\
L_1 \\
H_2 \\
H_1
\ea
\Longleftrightarrow
\ba{c} \CT_{Dec} \\
m_1 \\
m_2 \\
m_3 \\
l_3 \\
l_2 \\
l_1 \\
h_2 \\
h_1 
\ea
$}
\ee
Combining the two mappings \eqref{map1Usp6} and \eqref{map2Usp6}, we get the mapping between $\CT_0$ and $\CT_{Dec}$ 
\be \label{map3Usp6}
\scalebox{0.9}{$\ba{c}\CT_{1} \\
\tr(Q \, A^i \, Q) \\
\tr(Q \, A^j \, P) \\
\tr(A^k)
\ea
\quad \Longleftrightarrow \quad
\ba{c} \CT_{DEC} \\
m_{i+1}  \\
l_{3-j} \\
h_{N+1-k}\\
\ea
\qquad
\ba{l}\\
i= 0, 1, 2 \\
j= 0, 1, 2\\
k=2,3
\ea
$}
\ee

\subsection{Self-duality}
We already said that this theory is self-dual \cite{Csaki:1996eu}. Let us see now how we can use our $\CT_{Dec}$ frame to prove the self-duality. The strategy is to \emph{reconfine} the quiver tail. We notice that the left $Usp(2)$ has $6$ fundamentals, so we start confining from the left. The effect of this confinement is to kill the antisymmetric field $a_2$. In addition, the fields $\a_1$, $v_2$ and $h_2$ get a mass and we produce a Pfaffian superpotential as in \eqref{Switching2}. We get
\be \label{Usp6R1} \scalebox{0.84}{\bpic[node distance=2cm,gSUnode/.style={circle,red,draw,minimum size=8mm},gUSpnode/.style={circle,blue,draw,minimum size=8mm},fnode/.style={rectangle,draw,minimum size=8mm}]  
\node at (0,2) {$\cR_{1}:$};
\node[gUSpnode] (G2) at (3,0) {$4$};
\node[gUSpnode] (G3) at (6,0) {$6$};
\node[fnode,minimum width=1.4cm] (F1) at (9.5,0) {$7$};
\node[fnode] (F2) at (1.5,-2.8) {$1$};
\node[fnode,red] (F4) at (6,-2.8) {$1$};
\draw (G2) -- (G3);
\draw (G2) -- (F2);
\draw (G3) -- pic[pos=0.8,sloped,very thick]{arrow=latex reversed} (F1);
\draw (G2) -- (F4);
\draw (G3) -- (F4);
\draw (F2) -- (F4);
\draw[-*] (1.7,-3.2) -- (1.7,-4);
\draw[*-] (9.1,-1.4) -- pic[pos=0.8,sloped]{arrow} (9.1,-0.4);
\draw[-*] (1.3,-3.2) -- (1.3,-4);
\draw[*-] (10,-1.4) -- pic[pos=0.8,sloped]{arrow} (10,-0.4);
\draw[-*] (F4) -- (6,-4);
\draw (9.9,0.4) to[out=90,in=0] pic[pos=0.1,sloped]{arrow} (9.5,0.9) to[out=180,in=90] pic[pos=0.6,sloped,very thick]{arrow=latex reversed} (9.1,0.4);
\draw (9.5,1.3) node {$\vdots$};
\draw (10.1,0.4) to[out=90,in=0] pic[pos=0.2,sloped]{arrow} (9.5,1.7) to[out=180,in=90] pic[pos=0.6,sloped,very thick]{arrow=latex reversed} (8.9,0.4);
\node[right] at (-1.5,-5.2) {$\CW= m_1 \tr(b_2 b_2 b_2 q_3 q_3 b_2) + m_2 \tr(b_2 q_3 q_3 b_2) + m_3 q_3 q_3 + \a_2 \, \tr(b_2 b_2) + r_2 b_2 v_3 $};
\node[right] at (-1.5,-6.2) {$+ \, l_1 p_1 b_2 q_3 + l_3 v_3 q_3 + h_1 p_1 r_2 + l_2 p_1 b_2 b_2 b_2 q_3$};
\node at (9,-1.7) {$l_1$};
\node at (0.8,-4) {$l_1$};
\node at (9.6,-0.8) {$\dots$};
\node at (10.2,-1.7) {$l_3$};
\node at (2.2,-4) {$l_2$};
\node at (6.4,-4) {$l_3$};
\node at (4.7,-1.2) {$r_2$};
\node at (3.7,-1.7) {\scalebox{0.7}{$5-\frac{11}{2}R_A-7R_Q$}};
\node at (2,-1) {$p_1$};
\node at (0.8,-1.7) {\scalebox{0.7}{$-3+\frac{5}{2}R_A+7R_Q$}};
\node at (6.3,-1.2) {$v_3$};
\node at (7.1,-1.7) {\scalebox{0.7}{$4-5R_A-7R_Q$}};
\node at (4.5,0.4) {$b_2, \,$ \scalebox{0.7}{$\frac{1}{2} R_A$}};
\node at (3.7,-3.1) {$h_1, \,$ \scalebox{0.7}{$3 R_A$}};
\node at (9.8,2) {$m_1, \dots, m_3$};
\node at (7.5,-0.4) {$q_3, \,$ \scalebox{0.7}{$R_A+R_Q$}};
\epic} \ee
Then we can confine the $Usp(4)$ node. We will reach the self-dual frame of the original theory. Indeed, we produce a traceless antisymmetric field, $B$, for the $Usp(6)$ and the fields $h_1$, $\a_2$ and $v_3$ get a mass. The final quiver reads
\be \label{Usp6R2} \scalebox{0.85}{\bpic[node distance=2cm,gSUnode/.style={circle,red,draw,minimum size=8mm},gUSpnode/.style={circle,blue,draw,minimum size=8mm},fnode/.style={rectangle,draw,minimum size=8mm}]
\node at (-2.2,2.5) {$\cR_{2} \equiv \cR_{final}:$};
\node[gUSpnode] (G2) at (-1,0) {$6$};
\node[fnode,minimum width=1.4cm] (F1) at (1.8,0) {$7$};
\node[fnode,red] (F2) at (-1,-2.4) {$1$};
\draw (G2) -- pic[pos=0.7,sloped,very thick]{arrow=latex reversed} (F1); 
\draw (G2) -- (F2);
\draw[*-] (-0.7,-3.6) -- (-0.7,-2.8);
\node at (-1,-3.2) {$\dots$};
\draw[*-] (-1.3,-3.6) -- (-1.3,-2.8);
\draw[*-] (2.2,-1.4) -- pic[pos=0.8,sloped]{arrow} (2.2,-0.4);
\node at (1.8,-1) {$\dots$};
\draw[*-] (1.4,-1.4) -- pic[pos=0.8,sloped]{arrow} (1.4,-0.4);
\draw (2.2,0.4) to[out=90,in=0] pic[pos=0.1,sloped]{arrow} (1.8,0.9) to[out=180,in=90] pic[pos=0.6,sloped,very thick]{arrow=latex reversed} (1.4,0.4);
\node at (1.8,1.3) {$\vdots$};
\draw (2.4,0.4) to[out=90,in=0] pic[pos=0.2,sloped]{arrow} (1.8,1.7) to[out=180,in=90] pic[pos=0.6,sloped,very thick]{arrow=latex reversed} (1.2,0.4);
\draw (-0.7,0.3) to[out=90,in=0]  (-1,0.8) to[out=180,in=90] (-1.3,0.3);
\node[right] at (3.7,0) {$\CW = \, \sum_{j=1}^{3} m_j \, \tr(q_3 \, B^{3-j} \, q_3)$};
\node[right] at (4.4,-1) {$\, + \sum_{j=1}^{3} \, l_j \, \tr(p_{2} \, B^{j-1} \, q_3)$};
\node at (-1.3,-4) {$l_1$};
\node at (1.4,-1.7) {$l_1$};
\node at (-0.5,-4) {$l_{3}$};
\node at (2.4,-1.7) {$l_{3}$};
\node at (2.8,1.9) {$m_1,\dots,m_{3}$};
\node at (-1.3,-1.3) {$p_{2}$};
\node at (0.3,-0.3) {$q_{3}$};
\node at (-0.6,1) {$B$};
\epic} \ee
We can repackage the final result into a manifestly $SU(8)$ invariant way 
\be \label{Rfinal} \scalebox{0.85}{\bpic[node distance=2cm,gSUnode/.style={circle,red,draw,minimum size=8mm},gUSpnode/.style={circle,blue,draw,minimum size=8mm},fnode/.style={rectangle,draw,minimum size=8mm}]
\node at (-2.2,2.5) {$\cR_{final}:$};
\node[gUSpnode] (G2) at (-1,0) {$6$};
\node[fnode,minimum width=1.4cm] (F1) at (1.8,0) {$8$};
\draw (G2) -- pic[pos=0.7,sloped,very thick]{arrow=latex reversed} (F1); 
\draw (2.2,0.4) to[out=90,in=0] pic[pos=0.1,sloped]{arrow} (1.8,0.9) to[out=180,in=90] pic[pos=0.6,sloped,very thick]{arrow=latex reversed} (1.4,0.4);
\node at (1.8,1.3) {$\vdots$};
\draw (2.4,0.4) to[out=90,in=0] pic[pos=0.2,sloped]{arrow} (1.8,1.6) to[out=180,in=90] pic[pos=0.6,sloped,very thick]{arrow=latex reversed} (1.2,0.4);
\draw (-0.7,0.3) to[out=90,in=0]  (-1,0.8) to[out=180,in=90] (-1.3,0.3);
\node[right] at (3.5,0) {$\CW = \, \sum_{j=1}^{3} \mu_j \, \tr(\Qt \, B^{3-j} \, \Qt)$};
\node at (2.8,1.9) {$\mu_1,\dots, \mu_{3}$};
\node at (0.3,-0.4) {$\Qt$};
\node at (-0.6,1) {$B$};
\epic} \ee
Where we define
\be 
\mu_j =
\begin{pmatrix}
\bovermat{7}{\phantom{12}  m_j \phantom{12}} & \vdots & \bovermat{1}{l_{4-j}} \\
\dotfill & \vdots \dotfill & \dotfill \\
& \vdots &  0
\end{pmatrix}
\begin{aligned}
&\left. \begin{matrix}
\vphantom{O_N} \\
\\
\end{matrix} \! \!\right\}
7 \\
&\left. \begin{matrix}
\\
\end{matrix} \! \!\right\}
1\\
\end{aligned}, \qquad \Qt =
\begin{pmatrix}
\bovermat{7}{\phantom{12}  q_N \phantom{12}} & \vdots & \bovermat{1}{p_{3-1}}
\end{pmatrix}
\ee 
During this sequential reconfinement, from $\CT_{Dec}$ and $\cR_{final}$, the mapping is
\be \label{map4Usp6}
\scalebox{0.92}{$
\ba{c}\CT_{Dec} \\
m_i \\
l_i \\
h_2 \\
h_1 
\ea
\Longleftrightarrow
\ba{c} \cR_{1} \\
m_i \\
l_i \\
\tr(b_2 b_2)^2 \\
h_1
\ea
\Longleftrightarrow
\ba{c} \cR_{final} \\
m_i \\
l_i \\
\tr(B^2) \\
\tr(B^3) 
\ea
$}
\ee
Comparing with the mapping \eqref{map3Usp6}, we read the mapping for the self-duality

\be \label{map5Usp6}
\ba{c}\CT_{1} \\
\tr(Q \, A^i \, Q) \\
\tr(Q \, A^j \, P) \\
\tr(A^k) 
\ea
\quad \Longleftrightarrow \quad
\ba{c} \cR_{final} \\
m_{i+1}  \\
l_{3-j} \\
\tr(B^k)
\ea
\qquad
\ba{l}\\
i= 0, \dots, 2 \\
j= 0, \dots, 2\\
k= 2, \dots, 3
\ea
\ee
This is precisely the mapping given in \cite{Csaki:1996eu}. 

Notice that in the reconfinement the precise operator flipped by the singlet $H_2$ was crucial to obtain the duality with the correct amount of gauge singlets.

In the next section we generalize our discussion to arbitrary $N$ and $F$.

\section{Sequential deconfinement of $Usp(2N)$ with ${\tiny\ydiagram{1,1}} $ + $2F \, {\tiny\ydiagram{1}}$}\label{UspNF}

In this section we study the general case. The $Usp(2N)$ gauge theory with a traceless antisymmetric field $A$ and $2F$ complex chiral fields (the number of fundamentals should be even to avoid the global anomaly).

We proceed as in \cite{Benvenuti:2020gvy} to derive a chain of $2N$ dual frames consisting of quiver theories, with number of gauge nodes ranging from $1$ to $N$. Let us start with the first $\cT_0$ quiver
\be \label{UspT0} \scalebox{0.85}{\bpic[node distance=2cm,gSUnode/.style={circle,red,draw,minimum size=8mm},gUSpnode/.style={circle,blue,draw,minimum size=8mm},fnode/.style={rectangle,draw,minimum size=8mm}]  
\node at (-5,1) {$\CT_0:$};
\node[gUSpnode] (G1) at (-3,0) {$2N$};
\node[fnode] (F1) at (-0.5,0) {$2F$};
\draw (G1) -- pic[pos=0.7,sloped]{arrow} (F1);
\draw (-2.7,0.4) to[out=90,in=0]  (-3,0.8) to[out=180,in=90] (-3.3,0.4);
\node[right] at (-2.5,-1.2) {$ \CW = 0$};
\node at (-2.5,0.9) {$A$};
\node at (1,0) {$\equiv$};
\node[gUSpnode] (G2) at (2.5,0) {$2N$};
\node[fnode] (F2) at (5,0) {$2F-1$};
\node[fnode,red] (F3) at (2.5,-1.8) {$1$};
\draw (G2) -- pic[pos=0.7,sloped]{arrow} (F2) node[midway,above] {$Q$};
\draw (G2) -- (F3) node[midway,left] {$P$};
\draw (2.8,0.4) to[out=90,in=0]  (2.5,0.8) to[out=180,in=90] (2.2,0.4);
\node[right] at (2,-2.9) {$ \CW= 0$};
\node at (3.1,0.9) {$A$};
\epic} \ee  
\subsection{Deconfine and dualize: first step}\label{step1Sp}
We start by using the deconfinement \eqref{UspDeconfinement}, we obtain
\be \label{UspT0'} \scalebox{0.83}{\bpic[node distance=2cm,gSUnode/.style={circle,red,draw,minimum size=8mm},gUSpnode/.style={circle,blue,draw,minimum size=8mm},fnode/.style={rectangle,draw,minimum size=8mm}]  
\node at (6,2.2) {$\CT_{0'}:$};
\node[gUSpnode] (G3) at (7,0) {$2N$};
\node[gUSpnode] (G4) at (10.5,0) {$2N-2$};
\node[fnode] (F4) at (8.3,1.8) {$2F-1$};
\node[fnode] (F5) at (7,-2.5) {$1$};
\node[fnode,red] (F6) at (10.5,-2.5) {$1$};
\draw (G3) -- (G4);
\draw (G3) -- pic[pos=0.7,sloped]{arrow} (F4.south west);
\draw (G3) -- (F5) node[midway,left] {$p_1$};
\draw (G4) -- (F6) node[midway,right] {$p_2$};
\draw (F5) -- (F6) node[midway,below] {$h_1$};
\draw (F5.north east) -- (G4);
\node[right] at (12,-1.2) {$ \CW= p_1 c_1 d_1 + h_1 d_1 p_2 + \b_1 \, \tr(c_1 c_1) $};
\node at (8.7,0.3) {$c_1$};
\node at (8.5,-1) {$d_1$};
\node at (6.9,1.1) {$q_1$};
\epic} \ee
Then we dualize the $Usp(2N)$ node with \eqref{IPduality}. This step is the same as in \eqref{Usp6T1}.

\be \label{UspT1} \scalebox{0.84}{\bpic[node distance=2cm,gSUnode/.style={circle,red,draw,minimum size=8mm},gUSpnode/.style={circle,blue,draw,minimum size=8mm},fnode/.style={rectangle,draw,minimum size=8mm}]  
\node at (-2.5,2.3) {$\CT_{1}:$};
\node[gUSpnode] (G1) at (-1.5,0) {$2F-6$};
\node[gUSpnode] (G2) at (1.5,0)  {$2N-2$};
\node[fnode] (F1) at (4.5,0) {$2F-1$};
\node[fnode] (F2) at (-1.5,-2.2) {$1$};
\node[fnode,red] (F3) at (1.5,-2.2) {$1$};
\draw (G1) -- (G2);
\draw (G1) to[out=60,in=180] (1.3,2) to[out=0,in=120] pic[pos=0.8,sloped,very thick]{arrow=latex reversed} (F1.north west);
\draw (G1) -- (F2);
\draw (G2) -- (F3);
\draw (G2) -- pic[pos=0.7,sloped]{arrow} (F1); 
\draw (F2) -- (F3);
\draw[-*] (F2) -- (-1.5,-3.4);
\draw[*-] (4.5,-1.4) -- pic[pos=0.8,sloped]{arrow} (F1);
\draw (1.9,0.7) to[out=90,in=0]  (1.5,1.3) to[out=180,in=90] (1.1,0.7);
\draw (4.9,0.4) to[out=90,in=0] pic[pos=0.1,sloped]{arrow} (4.5,0.9) to[out=180,in=90] pic[pos=0.6,sloped,very thick]{arrow=latex reversed} (4.1,0.4);
\node[right] at (5.9,0) {$\CW= M_1 Q_1 Q_1 + Q_1 C_1 Q_2 + L_1 Q_1 V_1$};
\node[right] at (5.9,-1) {$+ \, \tr(C_1 \P C_1) + H_1 V_1 C_1 P_2$};
\node at (-2,-3.5) {$L_1$};
\node at (5,-1.5) {$L_1$};
\node at (-1.1,1.4) {$Q_1$};
\node at (3.2,-0.4) {$Q_2$};
\node at (-2,-1.2) {$V_1$};
\node at (2,-1.2) {$P_2$};
\node at (0,-0.4) {$C_1$};
\node at (0,-2.5) {$H_1$};
\node at (5.3,1) {$M_1$};
\node at (2.1,1.3) {$\P$};
\epic} \ee
The mapping of the chiral ring generators after this first step is the following
\be \label{mapUsp1}
\scalebox{0.9}{$
\ba{c}\CT_0 \\
Q Q \\
\tr(Q A^i Q) \\
\tr(Q A^j P) \\
\tr(Q A^{N-1} P) \\
\tr(A^m) \\
\tr(A^{N})
\ea
\Longleftrightarrow
\ba{c} \CT_{0'} \\
q_{1} q_{1} \\
\tr(q_{1} \, (c_1 c_1)^i  \, q_{1}) \\
\tr(q_{1} \, c_1 (c_1 c_1)^j \, p_{2}) \\
q_{1} \, p_1 \\
\tr((c_1 c_1)^m) \\
h_1
\ea
\Longleftrightarrow
\ba{c}\CT_1 \\
M_{1} \\
\tr(Q_{2} \,\P^{i-1} \, Q_{2}) \\
\tr(Q_{2} \, \P^{j} \, P_{2}) \\
L_1 \\
\tr(\P^m) \\
H_1
\ea
\qquad
\ba{l}\\
\\
i=1, \ldots, N-1 \\
j=0, \ldots, N-2 \\
\\
m=2,\ldots,N-1 \\
\\
\ea
$}
\ee
Now we iterate the procedure. We deconfine the traceless antisymmetric field, $\P$ and then we dualize. Let us write explicitly another step and then it will be enough to obtain the general story.  
\subsection{Second step} \label{step2Sp}
After the deconfinement we get
\be \label{UspT1'} \scalebox{0.84}{\bpic[node distance=2cm,gSUnode/.style={circle,red,draw,minimum size=8mm},gUSpnode/.style={circle,blue,draw,minimum size=8mm},fnode/.style={rectangle,draw,minimum size=8mm}]  
\node at (-2.5,3.3) {$\CT_{1'}:$};
\node[gUSpnode] (G1) at (-1.5,0) {$2F-6$};
\node[gUSpnode] (G2) at (1.5,0) {$2N-2$};
\node[gUSpnode] (G3) at (4.5,0) {$2N-4$};
\node[fnode] (F1) at (1.5,2.3) {$2F-1$};
\node[fnode] (F2) at (-1.5,-2.2) {$1$};
\node[fnode] (F3) at (1.5,-2.2) {$1$};
\node[fnode,red] (F4) at (4.5,-2.2) {$1$};
\node[right] at (6,0) {$\CW= m_1 q_1 q_1 + q_1 c_1 q_2 + l_1 v_1 q_1$};
\node[right] at (6,-1) {$+ \, \tr(c_1 c_2 c_2 c_1) + h_1 v_1 c_1 c_2 p_3$};
\node[right] at (6,-2) {$+ \, h_2 d_2 p_3 + p_2 c_2 d_2 + \b_2 \, \tr(c_2 c_2)$};
\draw (G1) -- (G2);
\draw (G2) -- (G3);
\draw (G1.north east) -- pic[pos=0.8,sloped,very thick]{arrow=latex reversed} (F1.south west);
\draw (G1) -- (F2);
\draw (G2) -- pic[pos=0.8,sloped]{arrow} (F1);
\draw (G2) -- (F3);
\draw (G3) -- (F4);
\draw (G3) -- (F3.north east);
\draw (F3) -- (F4);
\draw[-*] (F2) -- (-1.5,-3.4);
\draw[*-] (3.2,2.3) -- pic[pos=0.6,sloped,very thick]{arrow=latex reversed} (F1);
\draw (1.9,2.7) to[out=90,in=0] pic[pos=0.1,sloped]{arrow} (1.5,3.2) to[out=180,in=90] pic[pos=0.6,sloped,very thick]{arrow=latex reversed} (1.1,2.7);
\draw (F2.south east) to[out=-60,in=180] (1.5,-3.4) to[out=0,in=-120] (F4.south west);
\node at (-1.8,-3.6) {$l_1$};
\node at (3.6,2.2) {$l_1$};
\node at (3,-1.4) {$d_2$};
\node at (-0.4,1.4) {$q_1$};
\node at (1.1,1.4) {$q_2$};
\node at (-1.9,-1.2) {$v_1$};
\node at (1.2,-1.2) {$p_2$};
\node at (4.9,-1.2) {$p_3$};
\node at (0,0.4) {$c_1$};
\node at (3,0.4) {$c_2$};
\node at (0,-3) {$h_1$};
\node at (3,-2.5) {$h_2$};
\node at (2.4,3.2) {$m_1$};
\epic} \ee
Now we dualize the $Usp(2N-2)$ node. The fields $q_1$, $d_2$ and $\b_2$ get a mass. In addition $\tr(c_1 c_2 c_2 c_1)$ becomes a mass term therefore there will be no link between the $Usp(2F-6)$ and $Usp(2N-4)$ gauge group. After the integration of these massive fields and a rescaling we get
\be \label{UspT2} \scalebox{0.87}{\bpic[node distance=2cm,gSUnode/.style={circle,red,draw,minimum size=8mm},gUSpnode/.style={circle,blue,draw,minimum size=8mm},fnode/.style={rectangle,draw,minimum size=8mm}]  
\node at (-1,2.3) {$\CT_{2}:$};
\node[gUSpnode] (G1) at (0,0) {$2F-6$};
\node[gUSpnode] (G2) at (3,0) {$2(2F-6)$};
\node[gUSpnode] (G3) at (6,0) {$2N-4$};
\node[fnode] (F1) at (9,0) {$2F-1$};
\node[fnode] (F2) at (0,-2.4) {$1$};
\node[fnode] (F3) at (3,-2.4) {$1$};
\node[fnode,red] (F4) at (6,-2.4) {$1$};
\node[right] at (-3.5,-5.1) {$\CW= M_1 \tr(B_1 Q_2 Q_2 B_1) + M_2 Q_2 Q_2 + \tr(C_2 \P  C_2) + \tr(B_1 A_1 B_1) + \a_1 \tr(B_1 B_1)$};
\node[right] at (-3.5,-6.1) {$+ \, \tr(B_1 C_2 C_2 B_1) + R_1 B_1 V_2 + L_1 V_1 B_1 Q_2 + L_2 V_2 Q_2 + Q_2 C_2 Q_3 + H_1 V_1 B_1 C_2 P_3 + H_2 V_2 C_2 P_3$};
\draw (G1) -- (G2);
\draw (G2) -- (G3);
\draw (G2) to[out=60,in=180] (5.8,1.9) to[out=0,in=120] pic[pos=0.8,sloped,very thick]{arrow=latex reversed} (F1.north west);
\draw (G1) -- (F2);
\draw (G2) -- (F3);
\draw (G3) -- pic[pos=0.8,sloped]{arrow} (F1);
\draw (G3) -- (F4);
\draw (G1) -- (F3.north west);
\draw (F3) -- (F4);
\draw[-*] (F2) -- (0,-3.6);
\draw[*-] (8.7,-1.1) -- pic[pos=0.8,sloped]{arrow} (8.7,-0.4);
\draw[-*] (F3) -- (3,-3.6);
\draw[*-] (9.4,-1.1) -- pic[pos=0.8,sloped]{arrow} (9.4,-0.4);
\draw (9.4,0.4) to[out=90,in=0] pic[pos=0.1,sloped]{arrow} (9,0.9) to[out=180,in=90] pic[pos=0.6,sloped,very thick]{arrow=latex reversed} (8.6,0.4);
\draw (9.6,0.4) to[out=90,in=0] pic[pos=0.2,sloped]{arrow} (9,1.2) to[out=180,in=90] pic[pos=0.6,sloped,very thick]{arrow=latex reversed} (8.4,0.4);
\draw (0.4,0.7) to[out=90,in=0]  (0,1.3) to[out=180,in=90] (-0.4,0.7);
\draw (6.4,0.7) to[out=90,in=0]  (6,1.3) to[out=180,in=90] (5.6,0.7);
\draw (F2.south east) to[out=-60,in=180] (3,-4.2) to[out=0,in=-120] (F4.south west);
\node at (8.6,-1.4) {$L_1$};
\node at (0.4,-3.6) {$L_1$};
\node at (9.6,-1.4) {$L_2$};
\node at (3.4,-3.6) {$L_2$};
\node at (1.4,-1.4) {$R_1$};
\node at (-0.3,-1.3) {$V_1$};
\node at (3.3,-1.5) {$V_2$};
\node at (6.3,-1.3) {$P_3$};
\node at (1.4,0.3) {$B_1$};
\node at (4.7,-0.3) {$C_2$};
\node at (1.6,-3.6) {$H_1$};
\node at (4.5,-2.7) {$H_2$};
\node at (9.8,1.5) {$M_1,M_2$};
\node at (4.3,1.2) {$Q_2$};
\node at (7.5,-0.4) {$Q_3$};
\node at (6.6,1.2) {$\P$};
\node at (0.7,1.2) {$A_1$};
\epic} \ee
We recall that the antisymmetric field $A_1$ is traceles, as well as $\Phi$.

The mapping after this second step is given by\footnote{Using the table of R-charges of the fields given in appendix~\ref{RchargesTk} in Table \ref{table:1}, it is easy to check that the mapping is consitent.}
\be \label{mapUsp2}
\scalebox{0.9}{$
\ba{c}\CT_1 \\
M_{1} \\
Q_{2} \, Q_{2} \\
\tr(Q_{2} \, \P^{i-1} \, Q_{2}) \\
\tr(Q_{2} \, \P^{j} \, P_{2}) \\
\tr(Q_{2} \, \P^{N-2} \, P_{2}) \\
L_1 \\
\tr(\P^m) \\
\tr(\P^{N-1}) \\
H_1 \\
\ea
\Longleftrightarrow
\ba{c} \CT_{1'} \\
m_{1} \\
q_{2} \, q_{2} \\
\tr(q_{2} \, (c_2 c_2)^{i-1} \, q_{2}) \\
\tr(q_{2} \, c_2 (c_2 c_2)^j \, p_{3}) \\
q_{2} \, p_2 \\
l_1 \\
\tr((b_2 b_2)^m) \\
h_2 \\
h_1 \\
\ea
\Longleftrightarrow
\ba{c} \CT_2 \\
M_{1} \\
M_{2} \\
\tr(Q_{3} \, \P^{i-2} \, Q_{3}) \\
\tr(Q_{3} \, \P^j \, P_3) \\
L_2 \\
L_1 \\
\tr(\P^m) \\
H_2 \\
H_1 \\
\ea
\qquad
\ba{l} \\
\\
i=2, \dots, N-1 \\
j=0, \ldots, N-3 \\
\\
\\
m=2, \ldots, N-2 \\
\\
\ea
$}
\ee
\subsection{After k steps} \label{stepkSp}
After the iteration of $k$ steps, we get the following quiver
\be \label{UspTk} \scalebox{0.85}{\bpic[node distance=2cm,gSUnode/.style={circle,red,draw,minimum size=8mm},gUSpnode/.style={circle,blue,draw,minimum size=8mm},fnode/.style={rectangle,draw,minimum size=8mm}]
\node at (-5.5,2.8) {$\CT_{k}:$};
\node[gUSpnode] (G1) at (-5,0) {$2F-6$};
\node[gUSpnode] (G2) at (-2,0) {$2(2F-6)$};
\node (G3) at (0.2,0) {$\dots$};
\node[gUSpnode] (G4) at (2.6,0) {$k(2F-6)$};
\node[gUSpnode] (G5) at (5.6,0) {$2N-2k$};
\node[fnode] (F1) at (8.6,0) {$2F-1$};
\node[fnode] (F2) at (-5,-2.4) {$1$};
\node[fnode] (F3) at (-2,-2.4) {$1$};
\node[fnode] (F4) at (0.1,-2.4) {$1$};
\node at (1.4,-2.4) {$\dots$};
\node[fnode] (F5) at (2.6,-2.4) {$1$};
\node[fnode,red] (F6) at (5.6,-2.4) {$1$};
\node[right] at (-6.1,-6) {$\CW = \, \sum_{i=1}^{k} M_i \, \tr(B_i B_{i+1} \dots B_{k-1} \, Q_{k} \, Q_{k} \, B_{k-1} \dots B_{i+1} B_i) + \tr(C_{k} \,\P \, C_{k}) + \sum_{i=1}^{k-1} \a_i \tr(B_i B_i)$};
\node[right] at (-6.1,-7) {$+ \, \sum_{i=1}^{k-1} \tr(B_i \, A_i \, B_i) + \sum_{i=1}^{k-2} \tr(B_i \, A_{i+1} \, B_i) + \tr(B_{k-1} C_{k} C_{k} B_{k-1} ) + \sum_{i=1}^{k-1} R_i B_i V_{i+1}$};
\node[right] at (-6.1,-8) {$+ \, \sum_{i=1}^{k} L_i V_i B_i B_{i+1} \dots B_{k-1} Q_{k} + Q_{k} C_{k} Q_{k+1} + \sum_{i=1}^{k} H_i V_i B_i B_{i+1} \dots B_{k-1} C_{k} P_{k+1}$};
\draw (G1) -- (G2) -- (G3) -- (G4) -- (G5);
\draw (G1) -- (F2);
\draw (G1) -- (F3.north west);
\draw (G2) -- (F3);
\draw (G2) -- (F4.north west);
\draw (G3) -- (F5.north west);
\draw (G4) -- (F5);
\draw (G5) -- (F6);
\draw (G4) to[out=60,in=180] (5.4,2.1) to[out=0,in=120] pic[pos=0.8,sloped,very thick]{arrow=latex reversed} (F1.north west);
\draw (G5) -- pic[pos=0.8,sloped]{arrow} (F1); 
\draw (F5) -- (F6);
\draw[-*] (F2) -- (-5,-3.6);
\draw[-*] (F3) -- (-2,-3.6);
\draw[-*] (F5) -- (2.6,-3.6);
\draw[*-] (9,-1.5) -- pic[pos=0.8,sloped]{arrow} (9,-0.4);
\node at (8.6,-1) {$\dots$};
\draw[*-] (8.2,-1.5) -- pic[pos=0.8,sloped]{arrow} (8.2,-0.4);
\draw (9,0.4) to[out=90,in=0] pic[pos=0.1,sloped]{arrow} (8.6,0.9) to[out=180,in=90] pic[pos=0.6,sloped,very thick]{arrow=latex reversed} (8.2,0.4);
\node at (8.6,1.3) {$\vdots$};
\draw (9.2,0.4) to[out=90,in=0] pic[pos=0.2,sloped]{arrow} (8.6,1.6) to[out=180,in=90] pic[pos=0.6,sloped,very thick]{arrow=latex reversed} (8,0.4);
\draw (-4.6,0.7) to[out=90,in=0]  (-5,1.3) to[out=180,in=90] (-5.4,0.7);
\draw (-1.4,0.9) to[out=90,in=0]  (-2,1.6) to[out=180,in=90] (-2.6,0.9);
\draw (6.1,0.8) to[out=90,in=0]  (5.6,1.4) to[out=180,in=90] (5.1,0.8);
\draw (F2.south east) to[out=-50,in=180] (1.5,-5) to[out=0,in=-120] (F6.south east);
\draw (F3.south east) to[out=-60,in=180] (2.2,-4.5) to[out=0,in=-120] (F6.south);
\draw (F4.south) to[out=-60,in=180] (2.5,-4) to[out=0,in=-120] (F6.south west);
\node at (-4.9,-3.9) {$L_1$};
\node at (8,-1.8) {$L_1$};
\node at (-2.4,-3.5) {$L_2$};
\node at (3.1,-3.5) {$L_{k}$};
\node at (9.2,-1.8) {$L_{k}$};
\node at (-5.3,-1.3) {$V_1$};
\node at (-2.3,-1.5) {$V_2$};
\node at (3,-1.5) {$V_{k}$};
\node at (6.1,-1.4) {$P_{k+1}$};
\node at (-3.6,-1.4) {$R_1$};
\node at (-0.4,-1.3) {$R_2$};
\node at (1.1,-1.4) {$R_{k-1}$};
\node at (-3.6,0.4) {$B_1$};
\node at (-0.5,0.4) {$B_2$};
\node at (1,0.4) {$B_{k-1}$};
\node at (4.2,0.4) {$C_{k}$};
\node at (-3.7,-3.2) {$H_1$};
\node at (-0.8,-3.2) {$H_2$};
\node at (0.9,-3.2) {$H_3$};
\node at (4.2,-2.7) {$H_{k}$};
\node at (8.4,2) {$M_1,\dots,M_{k}$};
\node at (4,1.3) {$Q_{k}$};
\node at (7.2,-0.4) {$Q_{k+1}$};
\node at (-4.4,1.4) {$A_1$};
\node at (-1.1,1.7) {$A_2$};
\node at (6.2,1.4) {$\P$};
\epic} \ee
The last term in the superpotential should be taken with a grain of salt. Indeed as explained in the appendix~\ref{RchargesTk}, when $k$ is great enough some operators become degenerate with $V_i B_i B_{i+1} \dots B_{k-1} C_k P_{k+1}$ and then the superpotential should be modified. This is the \emph{degenerate holomorphic operator ambiguity} that we described in \eqref{Usp6T2}. Since it is a $k$-dependent statement, we decided to be cavalier when writing this term in \eqref{UspTk} and write the modified version in $\CT_{N-1}$. 

The mapping of the chiral ring generators is the following
\be \label{mapUsp3}
\scalebox{0.9}{$
\ba{c}\CT_{(k-1)'} \\
m_{1, \dots, k-1} \\
l_{1, \dots, k-1} \\
h_{1, \dots, k} \\
q_{k} \, q_{k} \\
\tr(q_{k} \, (c_{k} \, c_{k})^{i-k+1} \, q_{k}) \\
\tr(q_{k} \, c_{k} \, (c_{k} \, c_{k})^j \, p_{k+1}) \\
q_{k} \, p_{k} \\
\tr((c_k c_k)^m) 
\ea
\quad \Longleftrightarrow \quad 
\ba{c} \CT_{k} \\
M_{1, \dots, k-1} \\
L_{1, \dots, k-1} \\
H_{1, \dots, k} \\
M_{k} \\
\tr(Q_{k+1} \, \P^{i-k} \, Q_{k+1}) \\
\tr(Q_{k+1} \, \P^{j} \, P_{k+1}) \\
L_{k} \\
\tr(\P^m) 
\ea
\qquad
\ba{l}\\
\\
\\
\\
\\
i=k, \dots, N-1 \\
j=0, \dots, N-k-1 \\
\\
m=2,\dots, N-k 
\ea
$}
\ee

\subsection{After N-1 steps} \label{stepN-1Sp}

\be \label{UspTN-1} \scalebox{0.76}{\bpic[node distance=2cm,gSUnode/.style={circle,red,draw,minimum size=8mm},gUSpnode/.style={circle,blue,draw,minimum size=8mm},fnode/.style={rectangle,draw,minimum size=8mm}]
\node at (-5.3,2.3) {$\CT_{N-1}:$};
\node[gUSpnode] (G1) at (-5,0) {$2F-6$};
\node[gUSpnode] (G2) at (-2,0) {$2(2F-6)$};
\node (G3) at (0.4,0) {$\dots$};
\node[gUSpnode] (G4) at (3,0) {\scalebox{0.7}{$(N-1)(2F-6)$}};
\node[gUSpnode,minimum size=1cm] (G5) at (6,0) {$2$};
\node[fnode] (F1) at (8.6,0) {$2F-1$};
\node[fnode] (F2) at (-5,-2.4) {$1$};
\node[fnode] (F3) at (-2,-2.4) {$1$};
\node[fnode] (F4) at (0.4,-2.4) {$1$};
\node at (1.7,-2.4) {$\dots$};
\node[fnode] (F5) at (3,-2.4) {$1$};
\node[fnode,red] (F6) at (6,-2.4) {$1$};
\node[right] at (-5.9,-6) {$\CW = \, \sum_{i=1}^{N-1} M_i \, \tr(B_i B_{i+1} \dots B_{N-2} \, Q_{N-1} \, Q_{N-1} \, B_{N-2} \dots B_{i+1} B_i) + \sum_{i=1}^{N-2} \a_i \tr(B_i B_i)$};
\node[right] at (-5.9,-7) {$+ \, \sum_{i=1}^{N-2} \tr(B_i \, A_i \, B_i) + \sum_{i=1}^{N-3} \tr(B_i \, A_{i+1} \, B_i) +\tr(B_{N-2} C_{N-1} C_{N-1} B_{N-2})$};
\node[right] at (-5.9,-8) {$+ \, \sum_{i=1}^{N-2} R_i B_i V_{i+1} + \, \sum_{i=1}^{N-1} L_i V_i B_i B_{i+1} \dots B_{N-2} Q_{N-1} + Q_{N-1} C_{N-1} Q_{N}$};
\node[right] at (-5.9,-9) {$+ \, H_1 V_1 B_1 \dots B_{N-2} C_{N-1} P_N + \sum_{i=2}^{N-1} H_i V_1 B_1 \dots B_{N-1-i} R_{N-i} $};
\draw (G1) -- (G2) -- (G3) -- (G4) -- (G5);
\draw (G1) -- (F2);
\draw (G1) -- (F3.north west);
\draw (G2) -- (F3);
\draw (G2) -- (F4.north west);
\draw (G3) -- (F5.north west);
\draw (G4) -- (F5);
\draw (G5) -- (F6);
\draw (G4) to[out=60,in=180] (5.4,2.1) to[out=0,in=120] pic[pos=0.8,sloped,very thick]{arrow=latex reversed} (F1.north west);
\draw (G5) -- pic[pos=0.8,sloped]{arrow} (F1); 
\draw (F2) -- (F3);
\draw[-*] (F2.south west) -- (-5.4,-3.6);
\draw[-*] (F3) -- (-2,-3.6);
\draw[-*] (3.2,-2.8) -- (3.2,-3.6);
\draw[*-] (9,-1.5) -- pic[pos=0.8,sloped]{arrow} (9,-0.4);
\node at (8.6,-1) {$\dots$};
\draw[*-] (8.2,-1.5) -- pic[pos=0.8,sloped]{arrow} (8.2,-0.4);
\draw (9,0.4) to[out=90,in=0] pic[pos=0.1,sloped]{arrow} (8.6,0.9) to[out=180,in=90] pic[pos=0.6,sloped,very thick]{arrow=latex reversed} (8.2,0.4);
\node at (8.6,1.3) {$\vdots$};
\draw (9.2,0.4) to[out=90,in=0] pic[pos=0.2,sloped]{arrow} (8.6,1.6) to[out=180,in=90] pic[pos=0.6,sloped,very thick]{arrow=latex reversed} (8,0.4);
\draw (-4.6,0.7) to[out=90,in=0]  (-5,1.3) to[out=180,in=90] (-5.4,0.7);
\draw (-1.4,0.9) to[out=90,in=0]  (-2,1.6) to[out=180,in=90] (-2.6,0.9);
\draw (-5.3,-2.8) to[out=-60,in=180] (1.5,-5) to[out=0,in=-120] (F6.south);
\draw (-5,-2.8) to[out=-60,in=180] (-1,-4.5) to[out=0,in=-120] (F5.south);
\draw (-4.7,-2.8) to[out=-60,in=180] (-2.2,-4) to[out=0,in=-120] (F4.south);
\node at (-5.2,-3.9) {$L_1$};
\node at (8,-1.8) {$L_1$};
\node at (-1.6,-3.5) {$L_2$};
\node at (3.9,-3.5) {$L_{N-1}$};
\node at (9,-1.8) {$L_{N-1}$};
\node at (-5.3,-1.3) {$V_1$};
\node at (-2.3,-1.5) {$V_2$};
\node at (3.5,-1.6) {$V_{N-1}$};
\node at (6.4,-1.3) {$P_{N}$};
\node at (-3.6,-1.5) {$R_1$};
\node at (-0.2,-1.3) {$R_2$};
\node at (1.3,-1.4) {$R_{N-2}$};
\node at (-3.6,0.4) {$B_1$};
\node at (-0.5,0.4) {$B_2$};
\node at (1.2,0.4) {$B_{N-2}$};
\node at (4.9,0.4) {$C_{N-1}$};
\node at (-3.4,-2.8) {$H_{N-1}$};
\node at (0.8,-3.3) {$H_{N-2}$};
\node at (2.3,-4.2) {$H_{2}$};
\node at (5.8,-3.7) {$H_{1}$};
\node at (8.2,2) {$M_1,\dots,M_{N-1}$};
\node at (4.5,1.3) {$Q_{N-1}$};
\node at (7.2,-0.4) {$Q_{N}$};
\node at (-4.4,1.4) {$A_1$};
\node at (-1.1,1.7) {$A_2$};
\epic} \ee
The last term in the superpotential is the one after switching, see Appendix~\ref{RchargesTk}. Since the last node is $Usp(2)$, there is no antisymmetric and we can directly use \eqref{IPduality}.
\subsection{Final step and fully deconfined frame} \label{stepFinalSp}
\be \label{UspTDec} \scalebox{0.80}{\bpic[node distance=2cm,gSUnode/.style={circle,red,draw,minimum size=8mm},gUSpnode/.style={circle,blue,draw,minimum size=8mm},fnode/.style={rectangle,draw,minimum size=8mm}]
\node at (-5.3,2.2) {$\CT_{Dec}:$};
\node[gUSpnode] (G1) at (-5,0) {$2F-6$};
\node[gUSpnode] (G2) at (-2,0) {\scalebox{0.8}{$2(2F-6)$}};
\node (G3) at (0.1,0) {$\dots$};
\node[gUSpnode] (G4) at (2.5,0) {\scalebox{0.6}{$(N-1)(2F-6)$}};
\node[gUSpnode,minimum size=1cm] (G5) at (5.7,0) {\scalebox{0.8}{$N(2F-6)$}};
\node[fnode] (F1) at (8.7,0) {$2F-1$};
\node[fnode] (F2) at (-5,-2.4) {$1$};
\node[fnode] (F3) at (-2,-2.4) {$1$};
\node[fnode] (F4) at (0.1,-2.4) {$1$};
\node at (1.3,-2.4) {$\dots$};
\node[fnode] (F5) at (2.5,-2.4) {$1$};
\node[fnode,red] (F6) at (5.7,-2.4) {$1$};
\node[right] at (-5.7,-6) {$\CW = \, \sum_{i=1}^{N} m_i \, \tr(b_i b_{i+1} \dots b_{N-1} \, q_{N} \, q_{N} \, b_{N-1} \dots b_i b_{i+1} ) + \sum_{i=1}^{N-1} \a_i \tr(b_i b_i)$};
\node[right] at (-5.7,-7) {$+ \, \sum_{i=1}^{N-1} \tr(b_i \, a_i \, b_i) + \sum_{i=1}^{N-2} \tr(b_i \, a_{i+1} \, b_i) + \sum_{i=1}^{N-1} r_i b_i v_{i+1}$};
\node[right] at (-5.7,-8) {$+ \, \sum_{i=1}^{N} l_i v_i b_i b_{i+1} \dots b_{N-1} q_{N} + \sum_{i=1}^{N-1} h_i v_1 b_1 \dots b_{N-1-i} r_{N-i}$};
\draw (G1) -- (G2) -- (G3) -- (G4) -- (G5);
\draw (G1) -- (F2);
\draw (G1) -- (F3.north west);
\draw (G2) -- (F3);
\draw (G2) -- (F4.north west);
\draw (G3) -- (F5.north west);
\draw (G4) -- (F5);
\draw (G4) -- (F6.north west);
\draw (G5) -- (F6);
\draw (G5) -- pic[pos=0.8,sloped,very thick]{arrow=latex reversed} (F1); 
\draw (F2) -- (F3);
\draw[-*] (-5.4,-2.8) -- (-5.4,-3.6);
\draw[-*] (-1.7,-2.8) -- (-1.7,-3.6);
\draw[-*] (2.7,-2.8) -- (2.7,-3.6);
\draw[-*] (6.1,-2.8) -- (6.1,-3.6);
\draw[*-] (9.1,-1.5) -- pic[pos=0.8,sloped]{arrow} (9.1,-0.4);
\node at (8.7,-1) {$\dots$};
\draw[*-] (8.3,-1.5) -- pic[pos=0.8,sloped]{arrow} (8.3,-0.4);
\draw (9.1,0.4) to[out=90,in=0] pic[pos=0.1,sloped]{arrow} (8.7,0.9) to[out=180,in=90] pic[pos=0.6,sloped,very thick]{arrow=latex reversed} (8.3,0.4);
\node at (8.7,1.3) {$\vdots$};
\draw (9.3,0.4) to[out=90,in=0] pic[pos=0.2,sloped]{arrow} (8.7,1.6) to[out=180,in=90] pic[pos=0.6,sloped,very thick]{arrow=latex reversed} (8.1,0.4);
\draw (-4.6,0.7) to[out=90,in=0]  (-5,1.3) to[out=180,in=90] (-5.4,0.7);
\draw (-1.6,0.8) to[out=90,in=0]  (-2,1.4) to[out=180,in=90] (-2.4,0.8);
\draw (3,0.9) to[out=90,in=0]  (2.5,1.6) to[out=180,in=90] (2,0.9);
\draw (-5.3,-2.8) to[out=-60,in=180] (1.5,-5) to[out=0,in=-120] (5.9,-2.8);
\draw (-5,-2.8) to[out=-60,in=180] (-1,-4.5) to[out=0,in=-120] (2.5,-2.8);
\draw (-4.7,-2.8) to[out=-60,in=180] (-2,-4) to[out=0,in=-120] (0.1,-2.8);
\node at (-5.3,-3.9) {$l_1$};
\node at (8.2,-1.8) {$l_1$};
\node at (-1.3,-3.2) {$l_2$};
\node at (3.4,-3.5) {$l_{N-1}$};
\node at (6.6,-3.5) {$l_{N}$};
\node at (9.2,-1.8) {$l_{N}$};
\node at (-5.3,-1.3) {$v_1$};
\node at (-2.3,-1.4) {$v_2$};
\node at (3.1,-1.5) {$v_{N-1}$};
\node at (6.1,-1.4) {$v_{N}$};
\node at (-3.6,-1.4) {$r_1$};
\node at (-0.5,-1.4) {$r_2$};
\node at (0.9,-1.4) {$r_{N-2}$};
\node at (5,-1.3) {$r_{N-1}$};
\node at (-3.5,0.3) {$b_1$};
\node at (-0.6,0.3) {$b_2$};
\node at (0.9,0.3) {$b_{N-2}$};
\node at (4.2,0.3) {$b_{N-1}$};
\node at (-3.4,-2.7) {$h_{N-1}$};
\node at (0.2,-3.6) {$h_{N-2}$};
\node at (2,-4) {$h_{2}$};
\node at (5.4,-4.1) {$h_{1}$};
\node at (8.4,1.9) {$m_1,\dots,m_{N}$};
\node at (7.3,-0.3) {$q_{N}$};
\node at (-4.4,1.4) {$a_1$};
\node at (-1.3,1.4) {$a_2$};
\node at (3.5,1.4) {$a_{N-1}$};
\epic} \ee
We have obtained the fully \say{deconfined} frame. The mapping is
\be \label{map4}
\scalebox{0.88}{$\ba{c}\CT_{N-1} \\
M_{1, \dots, N-1} \\
L_{1, \dots, N-1} \\
H_{1, \dots, N-1} \\
Q_N \, Q_N \\
Q_N \, P_N
\ea
\quad \Longleftrightarrow \quad
\ba{c} \CT_{DEC} \\
m_{1, \dots, N-1}  \\
l_{1, \dots, N-1} \\
h_{1, \dots, N-1} \\
m_N \\
l_N
\ea
$}
\ee
Collecting all the mappings, we write the mapping from the $\CT_0$ frame to $\CT_{Dec}$\footnote{The mapping is consitent with the R-charges given in Table \ref{table:2}.}
\be \label{mapUsp5}
\scalebox{0.88}{$\ba{c}\CT_{0} \\
\tr(Q \, A^i \, Q) \\
\tr(Q \, A^j \, P) \\
\tr(A^k) 
\ea
\quad \Longleftrightarrow \quad
\ba{c} \CT_{DEC} \\
m_{i+1}  \\
l_{N-j} \\
h_{N+1-k}
\ea
\qquad
\ba{l}\\
i= 0, \dots, N-1 \\
j= 0, \dots, N-1\\
k= 2, \dots, N
\ea
$}
\ee
One interesting property of the \say{deconfined} frame is that all chiral ring generators are elementary gauge singlets.

\section{Reconfinement and self-duality for $Usp(2N)$ with ${\tiny\ydiagram{1,1}} $ + $8 \, {\tiny\ydiagram{1}}$}\label{reconf}
It was proposed in \cite{Csaki:1996eu} that in the special case of $F=4$ the theory studied in the previous section, $Usp(2N)$ with antisymmetric and $2F$ fundamentals, is self-dual modulo flips. In this section we use our $\CT_{Dec}$ frame \eqref{UspTDec} to prove this result. Let us rewrite it, specifying $F=4$:
\be \label{R0} \scalebox{0.8}{\bpic[node distance=2cm,gSUnode/.style={circle,red,draw,minimum size=8mm},gUSpnode/.style={circle,blue,draw,minimum size=8mm},fnode/.style={rectangle,draw,minimum size=8mm}]
\node at (-5.3,2) {$\cR_{0}:$};
\node[gUSpnode] (G1) at (-5,0) {$2$};
\node[gUSpnode] (G2) at (-2,0) {$4$};
\node (G3) at (0.1,0) {$\dots$};
\node[gUSpnode] (G4) at (2.5,0) {\scalebox{0.9}{$2(N-1)$}};
\node[gUSpnode,minimum size=1cm] (G5) at (5.7,0) {$2N$};
\node[fnode] (F1) at (8.7,0) {\hspace{0.3cm} $7\,\,\,\,\,\,$};
\node[fnode] (F2) at (-5,-2.4) {$1$};
\node[fnode] (F3) at (-2,-2.4) {$1$};
\node[fnode] (F4) at (0.1,-2.4) {$1$};
\node at (1.3,-2.4) {$\dots$};
\node[fnode] (F5) at (2.5,-2.4) {$1$};
\node[fnode,red] (F6) at (5.7,-2.4) {$1$};
\node[right] at (-5.7,-5.8) {$\CW = \, \sum_{i=1}^{N} m_i \, \tr(b_i b_{i+1} \dots b_{N-1} \, q_{N} \, q_{N} \, b_{N-1} \dots b_{i+1} b_i) + \sum_{i=1}^{N-1} \a_i \tr(b_i b_i)$};
\node[right] at (-5.7,-6.8) {$+ \, \sum_{i=2}^{N-1} \tr(b_i \, a_i \, b_i) + \sum_{i=1}^{N-2} \tr(b_i \, a_{i+1} \, b_i) + \sum_{i=1}^{N-1} r_i b_i v_{i+1}$};
\node[right] at (-5.7,-7.8) {$+ \, \sum_{i=1}^{N} l_i v_i b_i b_{i+1} \dots b_{N-1} q_{N} + \sum_{i=1}^{N-1} h_i v_1 b_1 \dots b_{N-1-i} r_{N-i}$};
\draw (G1) -- (G2);
\draw (G2) -- (G3) -- (G4) -- (G5);
\draw (G1) -- (F2);
\draw (G1) -- (F3.north west);
\draw (G2) -- (F3);
\draw (G2) -- (F4.north west);
\draw (G3) -- (F5.north west);
\draw (G4) -- (F5);
\draw (G4) -- (F6.north west);
\draw (G5) -- (F6);
\draw (G5) -- pic[pos=0.8,sloped,very thick]{arrow=latex reversed} (F1); 
\draw (F2) -- (F3);
\draw[-*] (-5.4,-2.8) -- (-5.4,-3.6);
\draw[-*] (-1.7,-2.8) -- (-1.7,-3.6);
\draw[-*] (2.7,-2.8) -- (2.7,-3.6);
\draw[-*] (6.1,-2.8) -- (6.1,-3.6);
\draw[*-] (9.1,-1.5) -- pic[pos=0.8,sloped]{arrow} (9.1,-0.4);
\node at (8.7,-1) {$\dots$};
\draw[*-] (8.3,-1.5) -- pic[pos=0.8,sloped]{arrow} (8.3,-0.4);
\draw (9.1,0.4) to[out=90,in=0] pic[pos=0.1,sloped]{arrow} (8.7,0.9) to[out=180,in=90] pic[pos=0.6,sloped,very thick]{arrow=latex reversed} (8.3,0.4);
\node at (8.6,1.3) {$\vdots$};
\draw (9.3,0.4) to[out=90,in=0] pic[pos=0.2,sloped]{arrow} (8.7,1.6) to[out=180,in=90] pic[pos=0.6,sloped,very thick]{arrow=latex reversed} (8.1,0.4);
\draw (-1.7,0.3) to[out=90,in=0]  (-2,0.8) to[out=180,in=90] (-2.3,0.3);
\draw (2.9,0.8) to[out=90,in=0]  (2.5,1.4) to[out=180,in=90] (2.1,0.8);
\draw (-5.3,-2.8) to[out=-60,in=180] (1.5,-5) to[out=0,in=-120] (5.9,-2.8);
\draw (-5,-2.8) to[out=-60,in=180] (-1,-4.5) to[out=0,in=-120] (2.5,-2.8);
\draw (-4.7,-2.8) to[out=-60,in=180] (-2,-4) to[out=0,in=-120] (0.1,-2.8);
\node at (-5.3,-3.9) {$l_1$};
\node at (8.2,-1.8) {$l_1$};
\node at (-1.3,-3.2) {$l_2$};
\node at (3.4,-3.5) {$l_{N-1}$};
\node at (6.6,-3.5) {$l_{N}$};
\node at (9.2,-1.8) {$l_{N}$};
\node at (-5.3,-1.3) {$v_1$};
\node at (-2.3,-1.4) {$v_2$};
\node at (3.1,-1.5) {$v_{N-1}$};
\node at (6.1,-1.4) {$v_{N}$};
\node at (-3.6,-1.4) {$r_1$};
\node at (-0.5,-1.4) {$r_2$};
\node at (0.9,-1.4) {$r_{N-2}$};
\node at (5,-1.3) {$r_{N-1}$};
\node at (-3.5,0.3) {$b_1$};
\node at (-0.9,0.3) {$b_2$};
\node at (1,0.3) {$b_{N-2}$};
\node at (4.3,0.3) {$b_{N-1}$};
\node at (-3.4,-2.7) {$h_{N-1}$};
\node at (0.2,-3.6) {$h_{N-2}$};
\node at (2,-4) {$h_{2}$};
\node at (5.4,-4.1) {$h_{1}$};
\node at (8.4,1.9) {$m_1,\dots,m_{N}$};
\node at (7.1,-0.3) {$q_{N}$};
\node at (-1.5,0.9) {$a_2$};
\node at (3.4,1.4) {$a_{N-1}$};
\epic} \ee
\be \label{mapSelfDuality1}
\scalebox{0.90}{$\ba{c}\CT_{0} \\
\tr(Q \, A^j \, Q) \\
\tr(Q \, A^j \, P) \\
\tr(A^i) 
\ea
\quad \Longleftrightarrow \quad
\ba{c} R_{0} \\
m_{j+1}  \\
l_{N-j} \\
h_{N+1-i}
\ea
\qquad
\ba{l}\\
j= 0,\dots, N-1 \\
j= 0,\dots, N-1\\
i= 2,\dots, N
\ea
$}
\ee
We see that the $Usp(2)$ gauge group is coupled to $4+1+1$ chiral fields in the fundamental representation and so it confines \eqref{IPconfining}. This step is similar as in \eqref{Usp6R1}. The confinement will give a mass to the traceless antisymmetric field $a_2$ as well as $v_2, \, \a_1$ and $h_{N-1}$. After integrating them out and computing the Pfaffian superpotential (See discussion below \eqref{Switching2}) we get
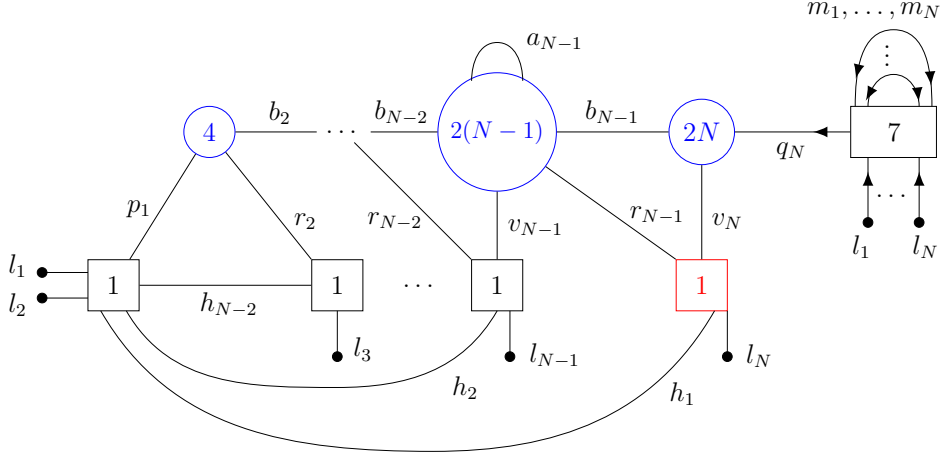
\be \label{R1} \scalebox{0.85}{\bpic[node distance=2cm,gSUnode/.style={circle,red,draw,minimum size=8mm},gUSpnode/.style={circle,blue,draw,minimum size=8mm},fnode/.style={rectangle,draw,minimum size=8mm}]
\node at (-5.1,2.3) {$\cR_{1}:$};
\node[gUSpnode] (G2) at (-2,0) {$4$};
\node (G3) at (0.1,0) {$\dots$};
\node[gUSpnode] (G4) at (2.5,0) {\scalebox{0.9}{$2(N-1)$}};
\node[gUSpnode,minimum size=1cm] (G5) at (5.7,0) {$2N$};
\node[fnode] (F1) at (8.7,0) {\hspace{0.3cm} $7\,\,\,\,\,\,$};
\node[fnode] (F2) at (-3.5,-2.4) {$1$};
\node[fnode] (F4) at (0,-2.4) {$1$};
\node at (1.3,-2.4) {$\dots$};
\node[fnode] (F5) at (2.5,-2.4) {$1$};
\node[fnode,red] (F6) at (5.7,-2.4) {$1$};
\node[right] at (-6,-6) {$\CW = \, m_1 \, \tr(b_2 b_2 b_2 \dots b_{N-1} \, q_N \, q_N \, b_{N-1} \dots b_2) + \sum_{i=2}^{N} m_i \, \tr(b_i b_{i+1} \dots b_{N-1} \, q_{N} \, q_{N} \, b_{N-1} \dots b_{i+1} b_i)$};
\node[right] at (-6,-7) {$+ \, \sum_{i=2}^{N-1} \a_i \tr(b_i b_i) + \sum_{i=3}^{N-1} \tr(b_i \, a_i \, b_i) + \sum_{i=2}^{N-2} \tr(b_i \, a_{i+1} \, b_i) + \sum_{i=2}^{N-1} r_i b_i v_{i+1}$};
\node[right] at (-6,-8) {$+ \, l_1 p_1 b_2 \dots b_{N-1} q_N + l_2 p_1 b_2 b_2 b_2 \dots b_{N-1} q_N + \sum_{i=3}^{N} l_i v_i b_i b_{i+1} \dots b_{N-1} q_{N} + \sum_{i=1}^{N-2} h_i p_1 b_2 \dots b_{N-1-i} r_{N-i}$};
\draw (G2) -- (G3);
\draw (G3) -- (G4) -- (G5);
\draw (G2) -- (F2);
\draw (G2) -- (F4.north west);
\draw (G3) -- (F5.north west);
\draw (G4) -- (F5);
\draw (G4) -- (F6.north west);
\draw (G5) -- (F6);
\draw (G5) -- pic[pos=0.8,sloped,very thick]{arrow=latex reversed} (F1); 
\draw (F2) -- (F4);
\draw[*-] (-4.7,-2.2) -- (-3.9,-2.2);
\draw[*-] (-4.7,-2.6) -- (-3.9,-2.6);
\draw[-*] (0,-2.8) -- (0,-3.6);
\draw[-*] (2.7,-2.8) -- (2.7,-3.6);
\draw[-*] (6.1,-2.8) -- (6.1,-3.6);
\draw[*-] (9.1,-1.5) -- pic[pos=0.8,sloped]{arrow} (9.1,-0.4);
\node at (8.7,-1) {$\dots$};
\draw[*-] (8.3,-1.5) -- pic[pos=0.8,sloped]{arrow} (8.3,-0.4);
\draw (9.1,0.4) to[out=90,in=0] pic[pos=0.1,sloped]{arrow} (8.7,0.9) to[out=180,in=90] pic[pos=0.6,sloped,very thick]{arrow=latex reversed} (8.3,0.4);
\node at (8.6,1.3) {$\vdots$};
\draw (9.3,0.4) to[out=90,in=0] pic[pos=0.2,sloped]{arrow} (8.7,1.6) to[out=180,in=90] pic[pos=0.6,sloped,very thick]{arrow=latex reversed} (8.1,0.4);
\draw (2.9,0.8) to[out=90,in=0]  (2.5,1.4) to[out=180,in=90] (2.1,0.8);
\draw (-3.7,-2.8) to[out=-60,in=180] (1.5,-5) to[out=0,in=-120] (5.9,-2.8);
\draw (-3.3,-2.8) to[out=-60,in=180] (0,-4.) to[out=0,in=-120] (2.5,-2.8);
\node at (-5,-2.1) {$l_1$};
\node at (8.2,-1.8) {$l_1$};
\node at (-5,-2.7) {$l_2$};
\node at (0.4,-3.4) {$l_3$};
\node at (3.4,-3.5) {$l_{N-1}$};
\node at (6.6,-3.5) {$l_{N}$};
\node at (9.2,-1.8) {$l_{N}$};
\node at (3.1,-1.5) {$v_{N-1}$};
\node at (6.1,-1.4) {$v_{N}$};
\node at (-0.5,-1.4) {$r_2$};
\node at (0.9,-1.4) {$r_{N-2}$};
\node at (5,-1.3) {$r_{N-1}$};
\node at (-0.9,0.3) {$b_2$};
\node at (1,0.3) {$b_{N-2}$};
\node at (4.3,0.3) {$b_{N-1}$};
\node at (-1.7,-2.7) {$h_{N-2}$};
\node at (2,-4) {$h_{2}$};
\node at (5.4,-4.1) {$h_{1}$};
\node at (8.4,1.9) {$m_1,\dots,m_{N}$};
\node at (-3.1,-1.2) {$p_{1}$};
\node at (7.1,-0.3) {$q_{N}$};
\node at (3.4,1.4) {$a_{N-1}$};
\epic} \ee
The mapping of the chiral ring generators is
\be \label{mapSelfDuality2}
\scalebox{0.90}{$\ba{c} R_{0} \\
m_j, \, l_j \\
h_i \\
h_{N-1}
\ea
\quad \Longleftrightarrow \quad
\ba{c} R_{1} \\
m_j, \, l_j   \\
h_i \\
\tr(b_2 b_2)^2
\ea
\quad
\ba{c} 
\text{it will stay like that} \\
i= 1,\dots, N-2 \\
\ea
$}
\ee
We can now iterate. Indeed the $Usp(4)$ group is coupled to $6+1+1$ fundamentals and so it also confines 

\be \label{R2} \scalebox{0.79}{\bpic[node distance=2cm,gSUnode/.style={circle,red,draw,minimum size=8mm},gUSpnode/.style={circle,blue,draw,minimum size=8mm},fnode/.style={rectangle,draw,minimum size=8mm}]
\node at (-5.1,2) {$\cR_{2}:$};
\node[gUSpnode] (G2) at (-2,0) {$6$};
\node (G3) at (0.1,0) {$\dots$};
\node[gUSpnode] (G4) at (2.5,0) {\scalebox{0.9}{$2(N-1)$}};
\node[gUSpnode,minimum size=1cm] (G5) at (5.7,0) {$2N$};
\node[fnode] (F1) at (8.7,0) {\hspace{0.3cm} $7\,\,\,\,\,\,$};
\node[fnode] (F2) at (-3.5,-2.4) {$1$};
\node[fnode] (F4) at (0,-2.4) {$1$};
\node at (1.3,-2.4) {$\dots$};
\node[fnode] (F5) at (2.5,-2.4) {$1$};
\node[fnode,red] (F6) at (5.7,-2.4) {$1$};
\node[right] at (-5.4,-6) {\scalebox{0.95}{$\CW = \, m_1 \, \tr([b_3 b_3]^3 b_4 \dots b_{N-1} \, q_N \, q_N \, b_{N-1} \dots b_4) + m_2 \, \tr([b_3 b_3]^2 b_4 \dots b_{N-1} \, q_N \, q_N \, b_{N-1} \dots  b_4) $}};
\node[right] at (-5.4,-7) {$+ \, \sum_{i=3}^{N} m_i \, \tr(b_i b_{i+1} \dots b_{N-1} \, q_{N} \, q_{N} \, b_{N-1} \dots b_{i+1} b_i) + \sum_{i=3}^{N-1} \a_{i} \tr(b_i b_i) + \sum_{i=4}^{N-1} \tr(b_i \, a_i \, b_i)$};
\node[right] at (-5.4,-8) {$+ \, \sum_{i=3}^{N-2} \tr(b_i \, a_{i+1} \, b_i) + \sum_{i=3}^{N-1} r_i b_i v_{i+1} + l_1 p_2 b_3 \dots b_{N-1} q_N + l_2 p_2 [b_3 b_3] b_3 \dots b_{N-1} q_N$};
\node[right] at (-5.4,-9) {$+ \, l_3 p_2 [b_3 b_3]^2 b_3 \dots b_{N-1} q_N + \sum_{i=4}^{N} l_i v_i b_i b_{i+1} \dots b_{N-1} q_{N} + \sum_{i=1}^{N-3} h_i p_2 b_3 \dots b_{N-1-i} r_{N-i}$};
\draw (G2) -- (G3) -- (G4) -- (G5);
\draw (G2) -- (F2);
\draw (G2) -- (F4.north west);
\draw (G3) -- (F5.north west);
\draw (G4) -- (F5);
\draw (G4) -- (F6.north west);
\draw (G5) -- (F6);
\draw (G5) -- pic[pos=0.8,sloped,very thick]{arrow=latex reversed} (F1); 
\draw (F2) -- (F4);
\draw[*-] (-4.7,-2.1) -- (-3.9,-2.1);
\node at (-4.2,-2.3) {$\vdots$};
\draw[*-] (-4.7,-2.7) -- (-3.9,-2.7);
\draw[-*] (0,-2.8) -- (0,-3.6);
\draw[-*] (2.7,-2.8) -- (2.7,-3.6);
\draw[-*] (6.1,-2.8) -- (6.1,-3.6);
\draw[*-] (9.1,-1.5) -- pic[pos=0.8,sloped]{arrow} (9.1,-0.4);
\node at (8.7,-1) {$\dots$};
\draw[*-] (8.3,-1.5) -- pic[pos=0.8,sloped]{arrow} (8.3,-0.4);
\draw (9.1,0.4) to[out=90,in=0] pic[pos=0.1,sloped]{arrow} (8.7,0.9) to[out=180,in=90] pic[pos=0.6,sloped,very thick]{arrow=latex reversed} (8.3,0.4);
\node at (8.6,1.3) {$\vdots$};
\draw (9.3,0.4) to[out=90,in=0] pic[pos=0.2,sloped]{arrow} (8.7,1.6) to[out=180,in=90] pic[pos=0.6,sloped,very thick]{arrow=latex reversed} (8.1,0.4);
\draw (2.9,0.8) to[out=90,in=0]  (2.5,1.4) to[out=180,in=90] (2.1,0.8);
\draw (-3.7,-2.8) to[out=-60,in=180] (1.5,-5) to[out=0,in=-120] (5.9,-2.8);
\draw (-3.3,-2.8) to[out=-60,in=180] (0,-4.) to[out=0,in=-120] (2.5,-2.8);
\node at (-5,-2.1) {$l_1$};
\node at (8.2,-1.8) {$l_1$};
\node at (-5,-2.7) {$l_3$};
\node at (0.4,-3.4) {$l_4$};
\node at (3.4,-3.5) {$l_{N-1}$};
\node at (6.6,-3.5) {$l_{N}$};
\node at (9.2,-1.8) {$l_{N}$};
\node at (3.1,-1.5) {$v_{N-1}$};
\node at (6.1,-1.4) {$v_{N}$};
\node at (-0.5,-1.4) {$r_3$};
\node at (0.9,-1.4) {$r_{N-2}$};
\node at (5,-1.3) {$r_{N-1}$};
\node at (-0.9,0.3) {$b_3$};
\node at (1,0.3) {$b_{N-2}$};
\node at (4.3,0.3) {$b_{N-1}$};
\node at (-1.7,-2.7) {$h_{N-3}$};
\node at (2,-4) {$h_{2}$};
\node at (5.4,-4.1) {$h_{1}$};
\node at (8.4,1.9) {$m_1,\dots,m_{N}$};
\node at (-3.1,-1.2) {$p_{2}$};
\node at (7.1,-0.3) {$q_{N}$};
\node at (3.4,1.4) {$a_{N-1}$};
\epic} \ee
We can iterate. After $k$ confinement, we get
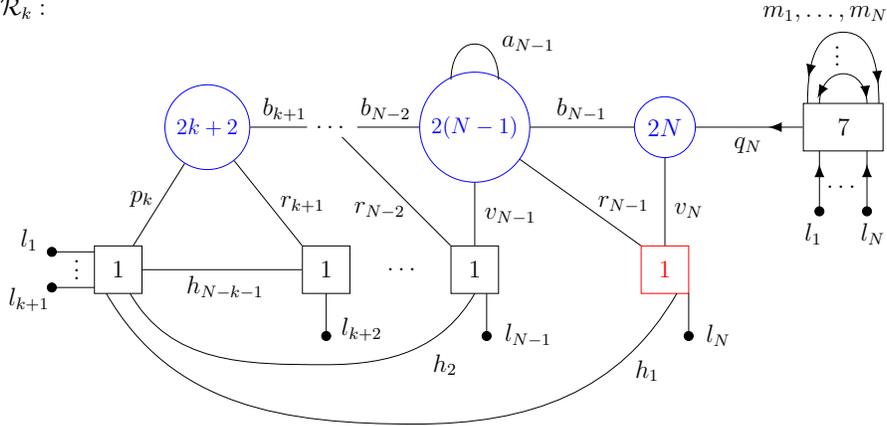
\be \label{Rk} \scalebox{0.79}{\bpic[node distance=2cm,gSUnode/.style={circle,red,draw,minimum size=8mm},gUSpnode/.style={circle,blue,draw,minimum size=8mm},fnode/.style={rectangle,draw,minimum size=8mm}]
\node at (-5.1,2) {$\cR_{k}:$};
\node[gUSpnode] (G2) at (-2,0) {\scalebox{0.9}{$2k+2$}};
\node (G3) at (0.1,0) {$\dots$};
\node[gUSpnode] (G4) at (2.5,0) {\scalebox{0.9}{$2(N-1)$}};
\node[gUSpnode,minimum size=1cm] (G5) at (5.7,0) {$2N$};
\node[fnode] (F1) at (8.7,0) {\hspace{0.3cm} $7\,\,\,\,\,\,$};
\node[fnode] (F2) at (-3.5,-2.4) {$1$};
\node[fnode] (F4) at (0,-2.4) {$1$};
\node at (1.3,-2.4) {$\dots$};
\node[fnode] (F5) at (2.5,-2.4) {$1$};
\node[fnode,red] (F6) at (5.7,-2.4) {$1$};
\node[right] at (-4.9,-6) {$\CW = \, \sum_{j=1}^{k} m_j \, \tr([b_{k+1} b_{k+1}]^{k+2-j} \, b_{k+2} \dots b_{N-1} \, q_N \, q_N \, b_{N-1} \dots b_{k+2})$};
\node[right] at (-4.9,-7) {$+ \, \sum_{i=k+1}^{N} m_i \, \tr(b_i b_{i+1} \dots b_{N-1} \, q_{N} \, q_{N} \, b_{N-1} \dots b_{i+1} b_i) + \sum_{i=k+1}^{N-1} \a_{i} \tr(b_i b_i)$};
\node[right] at (-4.9,-8) {$+ \, \sum_{i=k+2}^{N-1} \tr(b_i \, a_i \, b_i) + \sum_{i=k+1}^{N-2} \tr(b_i \, a_{i+1} \, b_i) + \sum_{i=k+1}^{N-1} r_i b_i v_{i+1}$};
\node[right] at (-4.9,-9) {$+ \, \sum_{j=1}^{k+1}  l_j \, p_k [b_{k+1} b_{k+1}]^{j-1} \, b_{k+1} \dots b_{N-1} q_N + \sum_{i=k+2}^{N} l_i v_i b_i b_{i+1} \dots b_{N-1} q_{N}$};
\node[right] at (-4.9,-10) {$+ \, \sum_{i=1}^{N-k-1} h_i p_k b_{k+1} \dots b_{N-1-i} r_{N-i}$};
\draw (G2) -- (G3) -- (G4) -- (G5);
\draw (G2) -- (F2);
\draw (G2) -- (F4.north west);
\draw (G3) -- (F5.north west);
\draw (G4) -- (F5);
\draw (G4) -- (F6.north west);
\draw (G5) -- (F6);
\draw (G5) -- pic[pos=0.8,sloped,very thick]{arrow=latex reversed} (F1); 
\draw (F2) -- (F4);
\draw[*-] (-4.7,-2.1) -- (-3.9,-2.1);
\node at (-4.2,-2.3) {$\vdots$};
\draw[*-] (-4.7,-2.7) -- (-3.9,-2.7);
\draw[-*] (0,-2.8) -- (0,-3.6);
\draw[-*] (2.7,-2.8) -- (2.7,-3.6);
\draw[-*] (6.1,-2.8) -- (6.1,-3.6);
\draw[*-] (9.1,-1.5) -- pic[pos=0.8,sloped]{arrow} (9.1,-0.4);
\node at (8.7,-1) {$\dots$};
\draw[*-] (8.3,-1.5) -- pic[pos=0.8,sloped]{arrow} (8.3,-0.4);
\draw (9.1,0.4) to[out=90,in=0] pic[pos=0.1,sloped]{arrow} (8.7,0.9) to[out=180,in=90] pic[pos=0.6,sloped,very thick]{arrow=latex reversed} (8.3,0.4);
\node at (8.6,1.3) {$\vdots$};
\draw (9.3,0.4) to[out=90,in=0] pic[pos=0.2,sloped]{arrow} (8.7,1.6) to[out=180,in=90] pic[pos=0.6,sloped,very thick]{arrow=latex reversed} (8.1,0.4);
\draw (2.9,0.8) to[out=90,in=0]  (2.5,1.4) to[out=180,in=90] (2.1,0.8);
\draw (-3.7,-2.8) to[out=-60,in=180] (1.5,-5) to[out=0,in=-120] (5.9,-2.8);
\draw (-3.3,-2.8) to[out=-60,in=180] (0,-4.) to[out=0,in=-120] (2.5,-2.8);
\node at (-5,-1.9) {$l_1$};
\node at (8.2,-1.8) {$l_1$};
\node at (-5,-2.9) {$l_{k+1}$};
\node at (0.6,-3.4) {$l_{k+2}$};
\node at (3.4,-3.5) {$l_{N-1}$};
\node at (6.6,-3.5) {$l_{N}$};
\node at (9.2,-1.8) {$l_{N}$};
\node at (3.1,-1.5) {$v_{N-1}$};
\node at (6.1,-1.4) {$v_{N}$};
\node at (-0.4,-1.3) {$r_{k+1}$};
\node at (0.9,-1.4) {$r_{N-2}$};
\node at (5,-1.3) {$r_{N-1}$};
\node at (-0.7,0.3) {$b_{k+1}$};
\node at (1,0.3) {$b_{N-2}$};
\node at (4.3,0.3) {$b_{N-1}$};
\node at (-1.7,-2.7) {$h_{N-k-1}$};
\node at (2,-4) {$h_{2}$};
\node at (5.4,-4.1) {$h_{1}$};
\node at (8.4,1.9) {$m_1,\dots,m_{N}$};
\node at (-3.1,-1.2) {$p_{k}$};
\node at (7.1,-0.3) {$q_{N}$};
\node at (3.4,1.4) {$a_{N-1}$};
\epic} \ee
The mapping between two successive reconfinement is
\be \label{mapSelfDuality4}
\scalebox{0.90}{$\ba{c} R_{k-1} \\
h_i \\
h_{N-k} \\
\tr(b_k b_k)^k \\
\vdots \\
\tr(b_k b_k)^2
\ea
\quad \Longleftrightarrow \quad
\ba{c} R_{k} \\
h_i \\
\tr(b_{k+1} b_{k+1})^{k+1} \\
\tr(b_{k+1} b_{k+1})^k \\
\vdots \\
\tr(b_{k+1} b_{k+1})^2
\ea
\qquad
\ba{c} 
i= 1,\dots, N-k-1 \\
\\
\\
\\
\ea
$}
\ee
For $k=N-2$, there won't be any antisymmetric field left. We get
\be \label{RN-2} \scalebox{0.83}{\bpic[node distance=2cm,gSUnode/.style={circle,red,draw,minimum size=8mm},gUSpnode/.style={circle,blue,draw,minimum size=8mm},fnode/.style={rectangle,draw,minimum size=8mm}]
\node at (-4,2) {$\cR_{N-2}:$};
\node[gUSpnode] (G2) at (-2,0) {\scalebox{0.9}{$2N-2$}};
\node[gUSpnode,minimum size=1cm] (G5) at (1.5,0) {$2N$};
\node[fnode,minimum width=1.6cm,minimum height=0.7cm] (F1) at (4,0) {$7$};
\node[fnode] (F2) at (-2,-2.4) {$1$};
\node[fnode,red] (F4) at (1.5,-2.4) {$1$};
\node[right] at (-6.2,-5.2) {$\CW = \, \sum_{j=1}^{N-2} m_j \, \tr([b_{N-1} b_{N-1}]^{N-j} \, q_N \, q_N) + m_{N-1} \, \tr(b_{N-1} \, q_N \, q_N \, b_{N-1}) + m_{N} q_{N} q_{N}$};
\node[right] at (-6.2,-6.2) {$+ \, \a_{N-1} \tr(b_{N-1} b_{N-1}) + r_{N-1} b_{N-1} v_{N} + \sum_{j=1}^{N-1} \, l_j \, p_{N-2} \, [b_{N-1} b_{N-1}]^{j-1} \, b_{N-1} \, q_N + l_{N} v_{N} q_{N}  + h_1 p_{N-2} r_{N-1}$};
\draw (G2) -- (G5);
\draw (G2) -- (F2);
\draw (G2) -- (F4.north west);
\draw (G5) -- (F4);
\draw (G5) -- pic[pos=0.7,sloped,very thick]{arrow=latex reversed} (F1); 
\draw (F2) -- (F4);
\draw[*-] (-1.7,-3.6) -- (-1.7,-2.8);
\node at (-2,-3.2) {$\dots$};
\draw[*-] (-2.3,-3.6) -- (-2.3,-2.8);
\draw[-*] (1.5,-2.8) -- (1.5,-3.6);
\draw[*-] (4.4,-1.5) -- pic[pos=0.8,sloped]{arrow} (4.4,-0.4);
\node at (4,-1) {$\dots$};
\draw[*-] (3.6,-1.5) -- pic[pos=0.8,sloped]{arrow} (3.6,-0.4);
\draw (4.4,0.4) to[out=90,in=0] pic[pos=0.1,sloped]{arrow} (4,0.9) to[out=180,in=90] pic[pos=0.6,sloped,very thick]{arrow=latex reversed} (3.6,0.4);
\node at (4,1.3) {$\vdots$};
\draw (4.6,0.4) to[out=90,in=0] pic[pos=0.2,sloped]{arrow} (4,1.6) to[out=180,in=90] pic[pos=0.6,sloped,very thick]{arrow=latex reversed} (3.4,0.4);
\node at (-2.4,-4) {$l_1$};
\node at (3.6,-1.8) {$l_1$};
\node at (-1.3,-4) {$l_{N-1}$};
\node at (1.6,-4) {$l_{N}$};
\node at (4.6,-1.8) {$l_{N}$};
\node at (1.9,-1.4) {$v_{N}$};
\node at (-0.5,-1.4) {$r_{N-1}$};
\node at (0,0.3) {$b_{N-1}$};
\node at (-0.2,-2.7) {$h_{1}$};
\node at (5,1.9) {$m_1,\dots,m_{N}$};
\node at (-2.6,-1.3) {$p_{N-2}$};
\node at (2.7,-0.3) {$q_{N}$};
\epic} \ee
Now we can do the last confinement with the $Usp(2N-2)$ group. It will produce the traceless antisymmetric field, $B$ for $Usp(2N)$ (the trace part is killed by the flipper $\a_{N-1}$). We get 
\be \label{RN-1} \scalebox{0.83}{\bpic[node distance=2cm,gSUnode/.style={circle,red,draw,minimum size=8mm},gUSpnode/.style={circle,blue,draw,minimum size=8mm},fnode/.style={rectangle,draw,minimum size=8mm}]
\node at (-2.2,2.2) {$\cR_{N-1}:$};
\node[gUSpnode] (G2) at (-1,0) {$2N$};
\node[fnode,minimum width=1.6cm,minimum height=0.7cm] (F1) at (1.8,0) {$7$};
\node[fnode,red] (F2) at (-1,-2.4) {$1$};
\node[right] at (3.7,0) {$\CW = \, \sum_{j=1}^{N} m_j \, \tr(q_N \, B^{N-j} \, q_N)$};
\node[right] at (4.4,-1) {$\, + \sum_{j=1}^{N} \, l_j \, \tr(p_{N-1} \, B^{j-1} \, q_N)$};
\draw (G2) -- pic[pos=0.7,sloped,very thick]{arrow=latex reversed} (F1); 
\draw (G2) -- (F2);
\draw[*-] (-0.7,-3.6) -- (-0.7,-2.8);
\node at (-1,-3.2) {$\dots$};
\draw[*-] (-1.3,-3.6) -- (-1.3,-2.8);
\draw[*-] (2.2,-1.5) -- pic[pos=0.8,sloped]{arrow} (2.2,-0.4);
\node at (1.8,-1) {$\dots$};
\draw[*-] (1.4,-1.5) -- pic[pos=0.8,sloped]{arrow} (1.4,-0.4);
\draw (2.2,0.4) to[out=90,in=0] pic[pos=0.1,sloped]{arrow} (1.8,0.9) to[out=180,in=90] pic[pos=0.6,sloped,very thick]{arrow=latex reversed} (1.4,0.4);
\node at (1.8,1.3) {$\vdots$};
\draw (2.4,0.4) to[out=90,in=0] pic[pos=0.2,sloped]{arrow} (1.8,1.6) to[out=180,in=90] pic[pos=0.6,sloped,very thick]{arrow=latex reversed} (1.2,0.4);
\draw (-0.7,0.4) to[out=90,in=0]  (-1,1) to[out=180,in=90] (-1.3,0.4);
\node at (-1.3,-4) {$l_1$};
\node at (1.4,-1.8) {$l_1$};
\node at (-0.5,-4) {$l_{N}$};
\node at (2.4,-1.8) {$l_{N}$};
\node at (2.8,1.9) {$m_1,\dots,m_{N}$};
\node at (-1.5,-1.3) {$p_{N-1}$};
\node at (0.3,-0.3) {$q_{N}$};
\node at (-0.5,1.1) {$B$};
\epic} \ee
The mapping for this last reconfinement is given by
\be \label{mapSelfDuality6} 
\ba{c} R_{N-2} \\
h_1 \\
\tr(b_{N-1} b_{N-1})^{N-1} \\
\tr(b_{N-1} b_{N-1})^{N-2} \\
\vdots \\
\tr(b_{N-1} b_{N-1})^2
\ea
\quad \Longleftrightarrow \quad
\ba{c} R_{N-1} \\
\tr(B^N) \\
\tr(B^{N-1}) \\
\tr(B^{N-2}) \\
\vdots \\
\tr(B^2)
\ea
\ee
We can repackage the last frame into a manifestly $SU(8)$ invariant way to obtain the final frame
\be \label{Rfinal} \scalebox{0.9}{\bpic[node distance=2cm,gSUnode/.style={circle,red,draw,minimum size=8mm},gUSpnode/.style={circle,blue,draw,minimum size=8mm},fnode/.style={rectangle,draw,minimum size=8mm}]
\node at (-2.2,2.5) {$\cR_{final}:$};
\node[gUSpnode] (G2) at (-1,0) {$2N$};
\node[fnode,minimum width=1.6cm,minimum height=0.7cm] (F1) at (1.8,0) {$8$};
\node[right] at (3.5,0) {$\CW = \, \sum_{j=1}^{N} \mu_j \, \tr(\Qt \, B^{N-j} \, \Qt)$};
\draw (G2) -- pic[pos=0.7,sloped,very thick]{arrow=latex reversed} (F1); 
\draw (2.2,0.4) to[out=90,in=0] pic[pos=0.1,sloped]{arrow} (1.8,0.9) to[out=180,in=90] pic[pos=0.6,sloped,very thick]{arrow=latex reversed} (1.4,0.4);
\node at (1.8,1.3) {$\vdots$};
\draw (2.4,0.4) to[out=90,in=0] pic[pos=0.2,sloped]{arrow} (1.8,1.6) to[out=180,in=90] pic[pos=0.6,sloped,very thick]{arrow=latex reversed} (1.2,0.4);
\draw (-0.7,0.4) to[out=90,in=0]  (-1,1) to[out=180,in=90] (-1.3,0.4);
\node at (2.8,1.9) {$\mu_1,\dots, \mu_{N}$};
\node at (0.3,-0.4) {$\Qt$};
\node at (-0.5,1.1) {$B$};
\epic} \ee
Where we define
\be 
\mu_j =
\begin{pmatrix}
\bovermat{7}{\phantom{12}  m_j \phantom{12}} & \vdots & \bovermat{1}{l_{N+1-j}} \\
\dotfill & \vdots \dotfill & \dotfill \\
& \vdots &  0
\end{pmatrix}
\begin{aligned}
&\left. \begin{matrix}
\vphantom{O_N} \\
\\
\end{matrix} \right\}
7 \\
&\left. \begin{matrix}
\\
\end{matrix} \right\}
1\\
\end{aligned}, \qquad \Qt =
\begin{pmatrix}
\bovermat{7}{\phantom{12}  q_N \phantom{12}} & \vdots & \bovermat{1}{p_{N-1}}
\end{pmatrix}
\ee 
Now combining the mappings, we see that the reconfinement procedure gives  
\be \label{mapSelfDuality7} 
\ba{c} R_{0} \\
h_i
\ea
\quad \Longleftrightarrow \quad
\ba{c} R_{final} \\
\tr(B^{N+1-i})
\ea
\qquad
\ba{c} \\
i= 1,\dots, N-1
\ea
\ee
Now if we compare the original frame $\CT_0$ and the last frame after reconfinement $\cR_{final}$ we see the self-duality and we obtain the following mapping
\be \label{mapSelfDuality8} 
\ba{c} T_{0} \\
\tr(Q \, A^j \, Q) \\
\tr(Q \, A^j \, P) \\
\tr(A^i) 
\ea
\quad \Longleftrightarrow \quad
\ba{c} R_{final} \\
m_{j+1}  \\
l_{N-j} \\
\tr(B^i)
\ea
\qquad
\ba{c} \\
j= 0,\dots, N-1 \\
j= 0,\dots, N-1\\
i= 2,\dots, N
\ea
\ee
Which is precisely the mapping proposed in \cite{Csaki:1996eu}.

\section{Reduction to $3d$ $\cN=2$ sequential deconfinement} \label{3d}
It is possible to reduce $4d$ $\cN=1$ theories on a circle, obtaining $3d$ $\cN=2$ theories. Generically, this steps introduces a superpotential term linear in the basic monopole operator (exceptions are, for instance, theories with $8$ supercharges). Once in $3d$, it is possible to turn on real mass deformations, that do not exist in $4d$. Starting from a $4d$ $Usp(2N)$ gauge theory, $3d$ real masses allow to flow to $Usp(2N)$ or $U(N)$ gauge groups with or without various types of monopole superpotentials. This process has been discussed in detail in the case without rank-$2$ matter \cite{Benini:2017dud}, and for the case of $2F=8$ \cite{Amariti:2018wht, Benvenuti:2018bav}. A brane interpretation has been found in \cite{Amariti:2017gsm}. Examples of $4d$ $\cN=1$ simplectic quivers reduced and deformed to $3d$ $\cN=2$ or $3d$ $\cN=4$ unitary quivers have been discussed in \cite{Pasquetti:2019tix, Pasquetti:2019hxf}, mostly from the superconformal index perspective. 

On the electric side, the story is as follows:
\be \label{3dredelectric} \scalebox{0.79}
{\bpic[node distance=2cm,gSUnode/.style={circle,red,draw,minimum size=8mm},gUSpnode/.style={circle,blue,draw,minimum size=8mm},fnode/.style={rectangle,draw,minimum size=8mm}]
\node[blue] at (-1,8) {$Usp(2N)$};
\node[fnode,minimum width=1.6cm,minimum height=0.7cm] (F1) at (1.8,8) {$2F+2$};
\node[right] at (3,8) {$\CW_{4d} = 0$};
\draw (-0.2,8) --  (F1); 
\draw (-0.7,8.4) to[out=90,in=0]  (-1,9) to[out=180,in=90] (-1.3,8.4);

\draw [red, line width=3pt] [-{Stealth[length=4mm, width=4mm]}] (0,7.2) -- (0,5.6);
\node[right] at (0.5, 6.4) {reduction on $S^1$};

\node[blue] at (-1,4) {$Usp(2N)$};
\node[fnode,minimum width=1.6cm,minimum height=0.7cm] (F2) at (1.8,4) {$2F+2$};
\node[right] at (3,4) {$\CW_{3d} = \M$}; \draw (-0.2,4) --  (F2); 
\draw (-0.7,4.4) to[out=90,in=0]  (-1,5) to[out=180,in=90] (-1.3,4.4);

\draw [red, line width=3pt] [-{Stealth[length=4mm, width=4mm]}] (1.5,3.2) -- (4,1.6);
\node[right] at (3, 2.4) {real masses $(+^{F+1},-^{F+1})$};

\draw [red, line width=3pt] [-{Stealth[length=4mm, width=4mm]}] (-0.5,3.2) -- (-3,1.6);
\node[left] at (-2.5, 2.4) {real masses $(0^{2F},+,-)$};

\node[blue] at (-6,0) {$Usp(2N)$};
\node[fnode,minimum width=1.6cm,minimum height=0.7cm] (F3) at (1.8-5,0) {$2F$};
\node[right] at (-2,0) {$\CW = 0$}; \draw (-5.2,0) -- (F3); 
\draw (-0.7-5,0.4) to[out=90,in=0]  (-6,1) to[out=180,in=90] (-1.3-5,0.4);

\node[red] at (3,0) {$U(N)$};
\node[fnode,minimum width=1.6cm,minimum height=0.7cm] (F4) at (5.8,0) {$(F+1,F+1)$};
\node[right] at (7.5,0) {$\CW = \M^+ + \M^-$}; \draw (3.5,0) -- (F4); 
\draw (4-0.7,0.4) to[out=90,in=0]  (4-1,1) to[out=180,in=90] (4-1.3,0.4);

\draw [red, line width=3pt] [-{Stealth[length=4mm, width=4mm]}] (4,7.2-8) -- (4,5.6-8);
\node[right] at (4.5, 6.4-8) {real masses $(0^F,+;0^F,-)$};

\node[red] at (3,-4) {$U(N)$};
\node[fnode,minimum width=1.6cm,minimum height=0.7cm] (F5) at (5.8,-4) {$(F,F)$};
\node[right] at (7.5,-4) {$\CW = \M^+ $}; \draw (3.5,-4) --  (F5); 
\draw (4-0.7,0.4-4) to[out=90,in=0]  (4-1,1-4) to[out=180,in=90] (4-1.3,0.4-4);

\draw [red, line width=3pt] [-{Stealth[length=4mm, width=4mm]}] (4,7.2-12) -- (4,5.6-12);
\node[right] at (4.5, 6.4-12) {real masses $(0^{F-1},+;0^{F-1},-)$};

\node[red] at (3,-8) {$U(N)$};
\node[fnode,minimum width=1.6cm,minimum height=0.7cm] (F6) at (5.8,-8) {$(F-1,F-1)$};
\node[right] at (7.5,-8) {$\CW = 0$}; \draw (3.5,-8) -- (F6); 
\draw (4-0.7,0.4-8) to[out=90,in=0]  (4-1,1-8) to[out=180,in=90] (4-1.3,0.4-8);
\epic} \ee
Where the rank-$2$ field is a traceless antisymmetric for $Usp(2N)$ and a traceless adjoint for $U(N)$. The monopoles $\M$, $\M^\pm$ are the  monopoles with minimal GNO charges. See \cite{Benini:2017dud} for more details.

We could also turn on different real masses, possibly leading to non-zero Chern-Simons terms as in \cite{Amariti:2018wht, Benvenuti:2018bav, Benvenuti:2020gvy}, but we refrain to do this in the present paper.

In the remaining of this section we perform the reduction and deformation of the fully deconfined dual, recovering the results found in \cite{Benvenuti:2020gvy} working in $3d$.

\subsubsection*{Reduction to the deconfined dual of $Usp(2N)$ with antisymmetric and $2F+2$ fundamentals, $\cW=\M$}
We put the $4d$ duality on a circle. On the electric side we get $3d$ $\cN=2$ $Usp(2N)$ with antisymmetric and $(2F+1)_Q+1_P$ fundamentals, $\cW=\M$, with global symmetry $SU(2F+2) \times U(1)$. 
On the magnetic side  we obtain the same quiver as in $4d$, \eqref{UspTDec}, with $F \rightarrow F+1$. The difference is that the superpontential now includes $N$ additional terms, linear in the monopole operators with GNO charges for a single gauge group, $\sum_{i=1}^N \M^{0^{i-1}, \bullet, 0^{N-i}}$: 
\be \label{DEC3d} \scalebox{0.83}{\bpic[node distance=2cm,gSUnode/.style={circle,red,draw,minimum size=8mm},gUSpnode/.style={circle,blue,draw,minimum size=8mm},fnode/.style={rectangle,draw,minimum size=8mm}]
\node[blue] (G1) at (-5,0) {\scalebox{0.8}{$Usp(2F\!-\!4)$}};
\node[blue] (G2) at (-2,0) {\scalebox{0.8}{$Usp(2(2F\!-\!4))$}};
\node (G3) at (0.1,0) {$\dots$};
\node[blue] (G4) at (2.5,0) {\scalebox{0.8}{$Usp((N\!-\!1)(2F\!-\!4))$}};
\node[blue] (G5) at (5.7,0) {\scalebox{0.8}{$Usp(N(2F\!-\!4))$}};
\node[fnode] (F1) at (8.7,0) {$2F+1$};
\node[fnode] (F2) at (-5,-2.4) {$1$};
\node[fnode] (F3) at (-2,-2.4) {$1$};
\node[fnode] (F4) at (0.1,-2.4) {$1$};
\node at (1.3,-2.4) {$\dots$};
\node[fnode] (F5) at (2.5,-2.4) {$1$};
\node[fnode,red] (F6) at (5.7,-2.4) {$1$};
\draw (G1) -- (G2) -- (G3) -- (G4) -- (G5);
\draw (G1) -- (F2);
\draw (G1) -- (F3.north west);
\draw (G2) -- (F3);
\draw (G2) -- (F4.north west);
\draw (G3) -- (F5.north west);
\draw (G4) -- (F5);
\draw (G4) -- (F6.north west);
\draw (G5) -- (F6);
\draw (G5) -- pic[pos=0.8,sloped,very thick]{arrow=latex reversed} (F1); 
\draw (F2) -- (F3);
\draw[-*] (-5.4,-2.8) -- (-5.4,-3.6);
\draw[-*] (-1.7,-2.8) -- (-1.7,-3.6);
\draw[-*] (2.7,-2.8) -- (2.7,-3.6);
\draw[-*] (6.1,-2.8) -- (6.1,-3.6);
\draw[*-] (9.1,-1.5) -- pic[pos=0.8,sloped]{arrow} (9.1,-0.4);
\node at (8.7,-1) {$\dots$};
\draw[*-] (8.3,-1.5) -- pic[pos=0.8,sloped]{arrow} (8.3,-0.4);
\draw (9.1,0.4) to[out=90,in=0] pic[pos=0.1,sloped]{arrow} (8.7,0.9) to[out=180,in=90] pic[pos=0.6,sloped,very thick]{arrow=latex reversed} (8.3,0.4);
\node at (8.7,1.3) {$\vdots$};
\draw (9.3,0.4) to[out=90,in=0] pic[pos=0.2,sloped]{arrow} (8.7,1.6) to[out=180,in=90] pic[pos=0.6,sloped,very thick]{arrow=latex reversed} (8.1,0.4);
\draw (-4.6,0.4) to[out=90,in=0]  (-5,1) to[out=180,in=90] (-5.4,0.4);
\draw (-1.6,0.4) to[out=90,in=0]  (-2,1) to[out=180,in=90] (-2.4,0.4);
\draw (3,0.4) to[out=90,in=0]  (2.5,1) to[out=180,in=90] (2,0.4);
\node at (-4.4,1) {$a_1$};\node at (-1.3,1) {$a_2$};\node at (3.5,1) {$a_{N-1}$};
\draw (-5.3,-2.8) to[out=-60,in=180] (1.5,-5) to[out=0,in=-120] (5.9,-2.8);
\draw (-5,-2.8) to[out=-60,in=180] (-1,-4.5) to[out=0,in=-120] (2.5,-2.8);
\draw (-4.7,-2.8) to[out=-60,in=180] (-2,-4) to[out=0,in=-120] (0.1,-2.8);
\node at (-5.3,-3.9) {$l_1$};
\node at (8.2,-1.8) {$l_1$};
\node at (-1.3,-3.2) {$l_2$};
\node at (3.4,-3.5) {$l_{N-1}$};
\node at (6.6,-3.5) {$l_{N}$};
\node at (9.2,-1.8) {$l_{N}$};
\node at (-5.3,-1.3) {$v_1$};
\node at (-2.3,-1.4) {$v_2$};
\node at (3.1,-1.5) {$v_{N-1}$};
\node at (6.1,-1.4) {$v_{N}$};
\node at (-3.6,-1.4) {$r_1$};
\node at (-0.5,-1.4) {$r_2$};
\node at (0.9,-1.4) {$r_{N-2}$};
\node at (5,-1.3) {$r_{N-1}$};
\node at (-3.5,0.3) {$b_1$};
\node at (-0.6,0.3) {$b_2$};
\node at (-3.4,-2.7) {$h_{N-1}$};
\node at (0.2,-3.6) {$h_{N-2}$};
\node at (2,-4) {$h_{2}$};
\node at (5.4,-4.1) {$h_{1}$};
\node at (8.4,1.9) {$m_1,\dots,m_{N}$};
\node at (7.3,-0.3) {$q_{N}$};
\node[right] at (-6.5,-6) {$\CW = \, \sum_{i=1}^{N} m_i \, \tr(b_i b_{i+1} \dots b_{N-1} \, q_{N} \, q_{N} \, b_{N-1} \dots b_i b_{i+1} ) + \sum_{i=1}^{N-1} \a_i \tr(b_i \, b_i) + \sum_{i=1}^{N-1} \tr(b_i \, a_i \, b_i)$};
\node[right] at (-6.5,-7) {$+ \, \sum_{i=1}^{N-2} \tr(b_i \, a_{i+1} \, b_i) + \sum_{i=1}^{N-1} r_i b_i v_{i+1} + \, \sum_{i=1}^{N} l_i v_i b_i b_{i+1} \dots b_{N-1} q_{N}+ \, \sum_{i=1}^{N-1} h_i v_1 b_1 \dots b_{N-1-i} r_{N-i}$};
\node[right] at (-6.5,-8) {$+ \, \sum_{i=1}^N \M^{0^{i-1}, \bullet, 0^{N-i}}$};
\epic} \ee
For convenience we reproduce also the mapping of the chiral ring generators:
\be\label{map3d1}
\scalebox{0.88}{$\ba{c}
\tr(Q \, A^i \, Q) \\
\tr(Q \, A^j \, P) \\
\tr(A^k) 
\ea
\quad \Longleftrightarrow \quad
\ba{c}
m_{i+1}  \\
l_{N-j} \\
h_{N+1-k}
\ea
\qquad
\ba{l}
i= 0, \dots, N-1 \\
j= 0, \dots, N-1\\
k= 2, \dots, N
\ea
$}
\ee
This is the same mapping as in $4d$, at this level there are no monopoles in the chiral ring, due to the presence of linear monopole terms in the superpotential.

\subsubsection*{Flow to the deconfined dual of $Usp(2N)$ with antisymmetric and $2F$ fundamentals, $\cW=0$}
We now  discuss what happens on the fully deconfined quiver \ref{DEC3d} upon turning on real masses.  We first turn on a real mass of the form $(0^{2F},+,-)$ (that is we are moving to the left in the diagram \ref{3dredelectric}), on the electric side $Q_{2F+1}$ and $P$ become massive. Notice that the rank of the global symmetry decrease by one unit. Accordingly the mesons $\tr(Q_I \, A^i \, Q_{2F+1})$ and $\tr(Q_I \, A^i \, P)$ have non-zero real mass, for $I=1,\ldots,2F$. Notice that $\tr(Q_{2F+1} \, A^i \, P)$ has zero total real mass. The electric theory becomes $3d$ $\cN=2$ $Usp(2N)$ with antisymmetric and $2F$ fundamentals, $\cW= 0$, with global symmetry $SU(2F) \times U(1) \times U(1)$.

It follows from the mapping \ref{map3d1} that the singlets $(l_{i})_I$ and $(m_{i} )_{I, 2F+1}$ become massive for $I=1,\ldots,2F$, while $(l_{i})_{2F+1}$ remain massless. The $l_j$'s and $(m_{i} )_{I, 2F+1}$'s become massive imply that also the elementary gauge variant fields $v_i, r_i$ and $(q_N)_{2F+1}$ become massive. In other words the saw structure in the fully deconfined quiver disappears, and we are left with
\be\label{DECusp3d} \scalebox{0.9}{\bpic[node distance=2cm,gSUnode/.style={circle,red,draw,minimum size=8mm},gUSpnode/.style={circle,blue,draw,minimum size=8mm},fnode/.style={rectangle,draw,minimum size=8mm}]
\node[blue] (G1) at (-5,0) {\scalebox{0.9}{$\!Usp(2F\!-\!4)\!$}};
\node[blue] (G2) at (-2,0) {\scalebox{0.9}{$\!Usp(2(2F\!-\!4))\!$}};
\node (G3) at (0.1,0) {$\!\dots\!$};
\node[blue] (G4) at (2.5,0) {\scalebox{0.9}{$\!Usp((N\!-\!1)(2F\!-\!4))\!$}};
\node[blue] (G5) at (6,0) {\scalebox{0.9}{$\!Usp(N(2F\!-\!4))\!$}};
\node[fnode] (F1) at (8.7,0) {$\,\,2\,F\,\,$};
\node[right] at (-5.7,-1.5) {$\CW = \, \sum_{i=1}^{N} m_i \, \tr(b_i b_{i+1} \dots b_{N-1} \, q \, q \, b_{N-1} \dots b_i b_{i+1} ) + \sum_{i=1}^{N-1} \a_i \tr(b_i \, b_i) + \sum_{i=1}^{N-1} \tr(b_i \, a_i \, b_i)$};
\node[right] at (-5.7,-2.5) {$+ \, \sum_{i=1}^{N-2} \tr(b_i \, a_{i+1} \, b_i)+ \, \sum_{i=1}^{N-1} h_i \M^{\bullet^{N-i}, 0^i} + \, \sum_{i=1}^{N}  (l_{i})_{2F+1} \M^{0^{i-1}, \bullet^{N-i+1}}$};
\draw (G1) -- (G2) -- (G3) -- (G4) -- (G5);
\draw (G5) -- pic[pos=0.8,sloped,very thick]{arrow=latex reversed} (F1); 
\draw (9.1,0.4) to[out=90,in=0] pic[pos=0.1,sloped]{arrow} (8.7,0.9) to[out=180,in=90] pic[pos=0.6,sloped,very thick]{arrow=latex reversed} (8.3,0.4);
\node at (8.7,1.3) {$\vdots$};
\draw (9.3,0.4) to[out=90,in=0] pic[pos=0.2,sloped]{arrow} (8.7,1.6) to[out=180,in=90] pic[pos=0.6,sloped,very thick]{arrow=latex reversed} (8.1,0.4);
\draw (-4.6,0.4) to[out=90,in=0]  (-5,1) to[out=180,in=90] (-5.4,0.4);
\draw (-1.6,0.4) to[out=90,in=0]  (-2,1) to[out=180,in=90] (-2.4,0.4);
\draw (3,0.4) to[out=90,in=0]  (2.5,1) to[out=180,in=90] (2,0.4);
\node at (-4.4,1) {$a_1$};\node at (-1.3,1) {$a_2$};\node at (3.5,1) {$a_{N-1}$};
\node at (-3.5,0.3) {$b_1$};\node at (-0.5,0.3) {$b_2$};
\node at (8.4,1.9) {$m_1,\dots,m_{N}$};
\node at (7.6,0.3) {$q$};
\epic} \ee
Notice that the linear monopole superpotential $\sum_{i=1}^N \M^{0^{i-1}, \bullet, 0^{N-i}}$ is lifted, while the massless gauge singlets $h_i$ and $(l_{i})_{2F+1}$ now flip monopole operators. (this is similar to the dimensional reduction of Seiberg and Intriligator-Pouliot dualities \cite{Aharony:2013dha}). These interactions are generated dynamically, one way to understand them is that such interactions are allowed by all global symmetries, and if they are not generated the gauge singlets would be free fields, which cannot be correct. 

Equation \eqref{DECusp3d} agrees with the results of section $2.4$ of \cite{Benvenuti:2020gvy} (modulo renaming $h_i \rightarrow \gamma_i$ and $(l_{i})_{2F+1} \rightarrow \sigma_i$), obtained by sequentually decofining in $3d$, using the deconfining duality $antisymm_{2N \times 2N} \leftrightarrow Usp(2N-2)-[2N], \cW= \gamma \M$. The only difference is the precise extended monopole flipped by $h_i$, the subtlety related to the degenerate holomorphic operators which can in principle be flipped by $h_i$ was not appreciated in  \cite{Benvenuti:2020gvy}. One can check that with the superpotential above, setting $F=3$, the tail reconfines appropriately and it is possible to derive the self-duality modulo flips of $3d$ $\cN=2$ $Usp(2N)$ with antisymmetric and $6$ fundamentals, $\cW= 0$, at each step one $h_i$ singlet is eaten, while the extended monopoles flipped by $(l_{i})_{2F+1}$ 'shorten' according to the rules of \cite{Benvenuti:2020wpc}.

\subsubsection*{Flow to the deconfined dual of $U(N)$ with adjoint and $F+1$ flavors, $\cW=\M^++\M^-$}
We now start from the $3d$ duality  $Usp(2N)$ with $2F+2$, $\cW=\M$ $\leftrightarrow$ \ref{DEC3d}, and turn on a real mass of the form $(+^{F+1},-^{F+1})$ (that is we are moving to the right in the diagram \ref{3dredelectric}). This type of real mass induces a Higgsing of the form $Usp(2 N_i) \rightarrow U(N_i)$ on both sides of the duality, the antisymmetric fields are replaced by adjoints and a pair of fundamentals is replaced by a fundamental plus an antifundamental. The Higgsing is induced by a vev of 'Coulomb branch type', that is, on both sides of the duality, we are going to a specific sublocus of the moduli space of vacua (inside the so called $\cN=2$ Coulomb branch) where there is the maximum amount of massless fields. Moreover, the monopole superpotentials $\M$'s are replaced by $(\M^++\M^-)$'s. See \cite{Benini:2017dud, Pasquetti:2019tix, Pasquetti:2019hxf} for more details.

On the electric side we flow to $U(N)$ with adjoint and $(F_Q+1_P,(F+1)_{\Qt})$ flavors and $\cW=\M^++\M^-$, with global symmetry $SU(F+1) \times SU(F+1) \times U(1)$\footnote{We split the global symmetry artificially into $SU(F+1)_{\Qt} \times SU(F)_Q \times U(1)_P \times U(1)_A$ to match the visible symmetries in the magnetic side.}. The rank of the global symmetry decreases by one unit, as it should. 

On the magnetic side, we end up with a fully deconfined quiver
\be \label{UmonDec} \scalebox{0.85}{\bpic[node distance=2cm,gSUnode/.style={circle,red,draw,minimum size=8mm},gUSpnode/.style={circle,blue,draw,minimum size=8mm},fnode/.style={rectangle,draw,minimum size=8mm}]
\node[red] (G1) at (-5,0) {\scalebox{0.8}{$U(F\!-\!2)$}};
\node[red] (G2) at (-2,0) {\scalebox{0.8}{$U(2(F\!-\!2))$}};
\node (G3) at (0.1,0) {$\dots$};
\node[red] (G4) at (2.5,0) {\scalebox{0.8}{$U((N\!-\!1)(F\!-\!2))$}};
\node[red,minimum size=1cm] (G5) at (5.7,0) {\scalebox{0.8}{$U(N(F\!-\!2))$}};
\node[fnode] (F1a) at (9.2,1) {$\,F\,$};
\node[fnode] (F1b) at (9.2,-1) {$\!F\!+\!1\!$};
\node[fnode] (F2) at (-5,-2.4) {$1$};
\node[fnode] (F3) at (-2,-2.4) {$1$};
\node[fnode] (F5) at (2.5,-2.4) {$1$};
\node[fnode] (F6) at (5.7,-2.4) {$1$};
\draw [latex-latex](G1) -- (G2);
\draw [latex-latex] (G2) -- (G3);
\draw [latex-latex] (G3) -- (G4);
\draw [latex-latex] (G4) -- (G5);
\draw [-latex] (G1) -- (F2);
\draw [latex-] (G1) -- (F3.north west);
\draw [-latex] (G2) -- (F3);
\draw [latex-] (G3) -- (F5.north west);
\draw [-latex] (G4) -- (F5);
\draw [latex-] (G4) -- (F6.north west);
\draw [-latex] (G5) -- (F6);
\draw [-latex] (G5) -- (F1a); 
\draw [latex-] (G5) -- (F1b); 
\node at (7.5,0.8) {$q$};
\node at (7.5,-0.8) {$\qt$};
\draw[-*] (-5.4,-2.8) -- (-5.4,-3.6);
\draw[-*] (-1.7,-2.8) -- (-1.7,-3.6);
\draw[-*] (2.7,-2.8) -- (2.7,-3.6);
\draw[-*] (6.1,-2.8) -- (6.1,-3.6);
\draw [-latex] (8.9,0.6) --  (8.9,-0.6);
\draw [-latex] (9.5,0.6) --  (9.5,-0.6);
\node at (9.9,0) {$M_i$};
\node at (9.2,0) {$\ldots$};
\draw [-latex] (9.6,-2.5) --  (9.6,-1.4);
\node at (9.3,-2) {$\dots$};
\draw [-latex]  (8.9,-2.5) -- (8.9,-1.4);
\draw (-4.6,0.4) to[out=90,in=0]  (-5,1) to[out=180,in=90] pic[pos=0.6,sloped,very thick]{arrow=latex reversed} (-5.4,0.4);
\draw (-1.6,0.4) to[out=90,in=0]  (-2,1) to[out=180,in=90] pic[pos=0.6,sloped,very thick]{arrow=latex reversed} (-2.4,0.4);
\draw (3,0.4) to[out=90,in=0]  (2.5,1) to[out=180,in=90] pic[pos=0.6,sloped,very thick]{arrow=latex reversed} (2,0.4);
\node at (9,-2.8) {$l_1$};\node at (9.8,-2.8) {$l_{N}$};
\node at (-5,-3.2) {$l_1$};
\node at (-1.3,-3.2) {$l_2$};
\node at (3.4,-3.2) {$l_{N-1}$};
\node at (6.6,-3.2) {$l_{N}$};
\node at (-5.3,-1.3) {$v_1$};
\node at (-2.3,-1.4) {$v_2$};
\node at (3.1,-1.5) {$v_{N-1}$};
\node at (6.1,-1.4) {$v_{N}$};
\node at (-3.6,-1.4) {$r_1$};
\node at (0.9,-1.4) {$r_{N-2}$};
\node at (5,-1.3) {$r_{N-1}$};
\node at (-3.5,0.3) {$b_1, \bt_1$};
\node at (-0.7,0.3) {$b_2, \bt_2$};
\node at (-4.4,1) {$a_1$};
\node at (-1.3,1) {$a_2$};
\node at (3.5,1) {$a_{N-1}$};
\node[right] at (-5.7,-4) {$\CW = \, \sum_{i=1}^{N} M_i \, \tr(b_i b_{i+1} \dots b_{N-1} \, q \, \qt \, \bt_{N-1} \dots \bt_i \bt_{i+1} ) + \sum_{i=1}^{N-1} \a_i \tr(b_i \, \bt_i) + \sum_{i=1}^{N-1} \tr(b_i \, a_i \, \bt_i)$};
\node[right] at (-5.2,-5) {$+ \, \sum_{i=1}^{N-2} \tr(b_i \, a_{i+1} \, \bt_i) + \sum_{i=1}^{N-1} tr(r_i b_i v_{i+1}) + \, \sum_{i=1}^{N} l_i tr(\qt \bt_{N-1} \dots \bt_{i+1} \bt_i v_i ) $};
\node[right] at (-5.2,-6) {$+ \, \sum_{i=1}^{N-1} h_i tr(   r_{N-i} \bt_{N-1-i} \dots \bt_1  v_1) + \sum_{i=1}^{N}  \left( \M^{0^{i-1}, +, 0^{N-i}} + \M^{0^{i-1}, -, 0^{N-i}} \right)$};
\epic} \ee
Where we did not draw the $h_i$ singlets\footnote{The double arrows $U(N_1) \leftrightarrow U(N_2)$ stand for a pair of bifundamentals with opposite orientation.}. The global symmetry of \eqref{UmonDec} is $SU(F+1) \times SU(F) \times U(1) \times U(1)$.

The mapping is\footnote{We have checked that the mapping is consistent by computing the R-charges of the operators (as a function of two variables coressponding to the two $U(1)$ symmetries) on both sides.}
\be\label{map3d2}
\ba{c}
\tr(\Qt_I \, A^i \, Q^J) \\
\tr(\Qt_I \, A^i \, P) \\
\tr(A^k) 
\ea
\quad \Longleftrightarrow \quad
\ba{c}
(M_{i+1})_{I}^J  \\
(l_{N-i})_I \\
h_{N+1-k}
\ea
\qquad
\ba{l}
i= 0, \dots, N-1 \\
i= 0, \dots, N-1 \\
k= 2, \dots, N
\ea
\ee
In the special case $F=3$, we can  use the confining duality for $U(N)$ with $(N+2, N+2)$ flavors and $\cW=\M^-+ \M^+$ \cite{Benini:2017dud}  and reconfine the tail in \ref{UmonDec}, deriving the self-duality modulo flips of $3d$ $\cN=2$ $U(N)$ with adjoint and $(4,4)$ fundamentals, $\cW= \M^++\M^-$, discussed in \cite{Benvenuti:2018bav}.

\subsubsection*{Flow to the deconfined dual of $U(N)$ with adjoint and $F$ flavors, $\cW=\M^+$}
We now turn on real masses $(0^F,+; 0^F,-)$ in the previous duality.

On the electric side one monopole superpotential is lifted and we flow to $U(N)$ with adjoint and $(F_Q, F_{\Qt})$ flavors and $\cW=\M^+$, with global symmetry $SU(F) \times SU(F) \times U(1) \times U(1)$. The rank of the global symmetry decreases by one unit, as expected. 

On the magnetic side, using \ref{map3d2}, the singlets $(l_{j})_I$ and $(M_{i})_{F+1}^J$ become massive. This in turn implies that $\qt_{F+1}$ gets a mass, together with the $v_i$'s and the $r_i$'s. Hence the saw disappears and we end up with

\be \label{UmoDec} \scalebox{0.9}{\bpic[node distance=2cm,gSUnode/.style={circle,red,draw,minimum size=8mm},gUSpnode/.style={circle,blue,draw,minimum size=8mm},fnode/.style={rectangle,draw,minimum size=8mm}]
\node[red] (G1) at (-5,0) {\scalebox{0.8}{$U(F\!-\!2)$}};
\node[red] (G2) at (-2,0) {\scalebox{0.8}{$U(2(F\!-\!2))$}};
\node (G3) at (0.1,0) {$\dots$};
\node[red] (G4) at (2.5,0) {\scalebox{0.8}{$U((N\!-\!1)(F\!-\!2))$}};
\node[red,minimum size=1cm] (G5) at (5.7,0) {\scalebox{0.8}{$U(N(F\!-\!2))$}};
\draw [latex-latex](G1) -- (G2);
\draw [latex-latex] (G2) -- (G3);
\draw [latex-latex] (G3) -- (G4);
\draw [latex-latex] (G4) -- (G5);
\node[fnode] (F1a) at (9.2,1) {$\,F\,$};
\node[fnode] (F1b) at (9.2,-1) {$\,F\,$};
\draw [-latex] (G5) -- (F1a); 
\draw [latex-] (G5) -- (F1b); 
\node at (7.5,0.8) {$q$};
\node at (7.5,-0.8) {$\qt$};
\draw [-latex] (8.9,0.6) --  (8.9,-0.6);
\draw [-latex] (9.5,0.6) --  (9.5,-0.6);
\node at (9.9,0) {$M_i$};
\node at (9.2,0) {$\ldots$};
\draw (-4.6,0.4) to[out=90,in=0]  (-5,1) to[out=180,in=90] pic[pos=0.6,sloped,very thick]{arrow=latex reversed} (-5.4,0.4);
\draw (-1.6,0.4) to[out=90,in=0]  (-2,1) to[out=180,in=90] pic[pos=0.6,sloped,very thick]{arrow=latex reversed} (-2.4,0.4);
\draw (3,0.4) to[out=90,in=0]  (2.5,1) to[out=180,in=90] pic[pos=0.6,sloped,very thick]{arrow=latex reversed} (2,0.4);
\node at (-3.5,0.3) {$b_1, \bt_1$};
\node at (-0.7,0.3) {$b_2, \bt_2$};
\node at (-4.4,1) {$a_1$};
\node at (-1.3,1) {$a_2$};
\node at (3.5,1) {$a_{N-1}$};
\node[right] at (-5.7,-2) {$\CW = \, \sum_{i=1}^{N} M_i \, \tr(b_i b_{i+1} \dots b_{N-1} \, q \, \qt \, \bt_{N-1} \dots \bt_i \bt_{i+1} ) + \sum_{i=1}^{N-1} \a_i \tr(b_i \, \bt_i) + \sum_{i=1}^{N-1} \tr(b_i \, a_i \, \bt_i)$};
\node[right] at (-5.2,-3) {$+ \, \sum_{i=1}^{N-2} \tr(b_i \, a_{i+1} \, \bt_i) + \sum_{i=1}^{N}  (l_i)_{2F+1}  \, \M^{0^{i-1}, +^{N-i+1}}   + \, \sum_{i=1}^{N-1} h_i   \M^{+^{N-i}, 0^{i}} + \, \sum_{i=1}^{N}  \M^{0^{i-1}, -, 0^{N-i}} $};
\epic} \ee
Notice that half of the linear monopole superpotential disappeared and that the massless gauge singlets $(l_i)_F$ and $h_i$ now flip monopole operators instead of mesons constructed with the saw.

In the special case $F=3$, we can use the confining duality for $U(N)$ with $(N+1,N+1)$ flavors and $\cW=\M^-+ h \M^+$ \cite{Benini:2017dud} to reconfine the tail in \ref{UmoDec}, deriving the self-duality modulo flips of $3d$ $\cN=2$ $U(N)$ with adjoint and $(3,3)$ fundamentals, $\cW= \M^+$ \cite{Benvenuti:2018bav}.

\subsubsection*{Flow to the deconfined dual of $U(N)$ with adjoint and $F-1$ flavors, $\cW=0$}
We now turn on real masses $(0^{F-1},+; 0^{F-1},-)$ in the previous duality.

On the electric side one monopole superpotential is lifted and we flow to $U(N)$ with adjoint and $F-1, F-1$ flavors and $\cW=0$, with global symmetry $SU(F-1) \times SU(F-1) \times U(1)^3$. Again, the rank of the global symmetry decreases by one unit.

On the magnetic side,

\be \label{UDec} \scalebox{0.85}{\bpic[node distance=2cm,gSUnode/.style={circle,red,draw,minimum size=8mm},gUSpnode/.style={circle,blue,draw,minimum size=8mm},fnode/.style={rectangle,draw,minimum size=8mm}]
\node[red] (G1) at (-5,0) {\scalebox{0.8}{$U(F\!-\!2)$}};
\node[red] (G2) at (-2,0) {\scalebox{0.8}{$U(2(F\!-\!2))$}};
\node (G3) at (0.1,0) {$\dots$};
\node[red] (G4) at (2.5,0) {\scalebox{0.8}{$U((N\!-\!1)(F\!-\!2))$}};
\node[red,minimum size=1cm] (G5) at (5.7,0) {\scalebox{0.8}{$U(N(F\!-\!2))$}};
\node[fnode] (F1a) at (9.2,1) {$\,F\!-\!1\,$};
\node[fnode] (F1b) at (9.2,-1) {$\,F\!-\!1\,$};
\draw [-latex] (G5) -- (F1a); 
\draw [latex-] (G5) -- (F1b); 
\node at (7.5,0.8) {$q$};
\node at (7.5,-0.8) {$\qt$};
\draw [-latex] (8.9,0.6) --  (8.9,-0.6);
\draw [-latex] (9.5,0.6) --  (9.5,-0.6);
\node at (9.9,0) {$M_i$};
\node at (9.2,0) {$\ldots$};
\draw [latex-latex](G1) -- (G2);
\draw [latex-latex] (G2) -- (G3);
\draw [latex-latex] (G3) -- (G4);
\draw [latex-latex] (G4) -- (G5);
\draw (-4.6,0.4) to[out=90,in=0]  (-5,1) to[out=180,in=90] pic[pos=0.6,sloped,very thick]{arrow=latex reversed} (-5.4,0.4);
\draw (-1.6,0.4) to[out=90,in=0]  (-2,1) to[out=180,in=90] pic[pos=0.6,sloped,very thick]{arrow=latex reversed} (-2.4,0.4);
\draw (3,0.4) to[out=90,in=0]  (2.5,1) to[out=180,in=90] pic[pos=0.6,sloped,very thick]{arrow=latex reversed} (2,0.4);
\node at (-3.5,0.3) {$b_1, \bt_1$};
\node at (-0.7,0.3) {$b_2, \bt_2$};
\node at (-4.4,1) {$a_1$};
\node at (-1.3,1) {$a_2$};
\node at (3.5,1) {$a_{N-1}$};
\node[right] at (-6.7,-2) {$\CW = \, \sum_{i=1}^{N} M_i \, \tr(b_i b_{i+1} \dots b_{N-1} \, q \, \qt \, \bt_{N-1} \dots \bt_i \bt_{i+1} ) + \sum_{i=1}^{N-1} \a_i \tr(b_i \, \bt_i) + \sum_{i=1}^{N-1} \tr(b_i \, a_i \, \bt_i)$};
\node[right] at (-6.2,-3) {$+ \, \sum_{i=1}^{N-2} \tr(b_i \, a_{i+1} \, \bt_i) + \sum_{i=1}^{N} \left( (l_i)_{2F+1}  \, \M^{0^{i-1}, +^{N-i+1}} +  (M_i)_{F,F}  \, \M^{0^{i-1}, -^{N-i+1}} \right)$};
\node[right] at (-6.2,-4) {$+ \, \sum_{i=1}^{N-1}  h_i   \, \M^{+^i, 0^{N-i}} + \sum_{i=1}^{N-1} \M^{0^{i-1}, -, 0^{N-i}}$};
\epic} \ee

The result \ref{UDec} agrees with section $3.2$ of \cite{Benvenuti:2020gvy} (modulo renaming $h_i \rightarrow \gamma_i$, $(l_{i})_{2F+1} \rightarrow \sigma^+_i$, $(M_{i})_{F,F} \rightarrow \sigma_i^-$ and $F \rightarrow F+1$). Also here, the  difference is the precise extended monopole flipped by $h_i$, the subtlety related to the degenerate holomorphic operators which can in principle be flipped by $h_i$ was not appreciated in  \cite{Benvenuti:2020gvy}. In the special case $F=3$, we can use the confining duality for $U(N)$ with $(N+1,N+1)$ flavors and $\cW=\M^-+ h \M^+$ to reconfine the tail in \ref{UDec} and derive the self-duality modulo flips of $3d$ $\cN=2$ $U(N)$ with adjoint and $(2,2)$ fundamentals, $\cW= 0$, discussed in \cite{Benvenuti:2018bav}.\footnote{At each reconfining step, one linear monopole term disappears, one $h_i$ is eaten, while the $(M_i)_{I}^J$'s ($I,J=1,2$) and the two towers $(l_i)_{2F+1}$ and $ (M_i)_{F,F}$ survive. The self-duality reads
\be U(N) + \Phi + (2_Q, 2_{\Qt}), \cW=0. \Leftrightarrow U(N) + \phi + (2_q, 2_{\qt}), \cW= \sum_i \left( M_i tr(\qt \phi^i q) + (l_i)_{2F+1} \M^+_{\phi^i} + (M_i)_{F,F} \M^-_{\phi^i} \right) .\ee}

\acknowledgments
We are grateful to Simone Giacomelli, Sara Pasquetti and Shlomo Razamat for useful conversations.\\
Stephane Bajeot is partially supported by the INFN Research Project STEFI. Sergio Benvenuti is partially supported by the INFN Research Project GAST.

\appendix
\section{R-charges and degenerate holomorphic operator ambiguity in generic frame $\CT_k$} \label{RchargesTk}

We start by writing the R-charges of the fields in the generic frame $\CT_k$ \eqref{UspTk}.
\begin{table}[H]
\centering
\begin{tabular}{|c|c|}
\hline
Fields & R-charges in $\CT_{k}$ \\
\hline
$V_i$ & $R_{V_i} = 5 - 2F + (N - \frac{5}{2} + \frac{3i}{2})R_A + (2F-1)R_Q$ \\
\hline
$L_i$ & $R_{L_i} = 2F-4 - (N-2+i)R_A - (2F-2)R_Q$ \\
\hline
$M_i$ & $R_{M_i} = 2R_Q + (i-1)R_A$ \\
\hline
$H_i$ & $(N-i+1)R_A$ \\
\hline
$R_j$ & $R_{R_j} = 2F-3 - (N + \frac{3j}{2} - \frac{1}{2})R_A - (2F-1)R_Q$ \\
\hline
$B_j$ & $R_{B_j} = \frac{1}{2}R_A$ \\
\hline
$A_j$ & $R_{A_j} = 2 - R_A$ \\
\hline
$C_k$ & $R_{C_k} = 1 - \frac{1}{2}R_A$ \\
\hline
$Q_k$ & $R_{Q_k} = 1 + \frac{1}{2}(1-k)R_A$ \\
\hline
$Q_{k+1}$ & $R_{Q_{k+1}} = R_Q + \frac{k}{2}R_A$ \\
\hline
$P_{k+1}$ & $R_{F_{k+1}} = 2F-4 - (2N - 2 + \frac{k}{2})R_A - (2F-1)R_Q$ \\
\hline
$\P$ & $R_{\P} = R_A$ \\
\hline
\end{tabular}
\caption{R-charges in the frame $\CT_k$ with $k=1, \dots, N-1$, $i=1, \dots, k$ and $j=1, \dots, k-1$.}
\label{table:1}
\end{table}

Now we want to find the degenerate operators that can couple to $H_i$, as we did in \eqref{Usp6T2} in the case of $Usp(6)$. In order to so, we look at the R-charges of the operator $V_i B_i \dots B_{k-1} C_k P_{k+1}$ which is the natural candidate to be coupled to $H_i$. We find
\be \label{RchargesDegOp}
R(V_i B_i \dots B_{k-1} C_k P_{k+1}) = 2- (N-i+1) R_A
\ee
The potential degenerate operators should be a singlet under the non-abelian global symmetry and should have the same R-charges \eqref{RchargesDegOp}. We can build the degenerate operators from the fields $V_a$, $B_b$ and $R_c$. Indeed, we start from $V_m$, then we put some $B_b$ and end with $R_n$ ($n \ge m$)\footnote{We cannot end with $V_{j+1}$ because the F-term equation of $R_j$ sets the combination $B_j V_{j+1}$ to $0$.}. The form of these operators is then: $V_m B_m \dots B_{n-1} R_n$. The number of fields $B$ is $n-m$. The R-charge is
\be 
R_{V_m} + (n-m) R_{B_b} + R_{R_n} = 2 - (2 + (n-m)) R_A
\ee 
If we compare with \eqref{RchargesDegOp}, we find the following condition
\be \label{DegOpConstraint}
n-m = N-i-1
\ee
with $1 \le m \le n \le k-1$, $i=1, \dots, k$ and $k=1, \dots, N-1$.

\noindent We can already make two remarks:
\begin{itemize}
\item For $i=1$, the constraint becomes $n-m = N-2$ but the maximal value of $n-m$ is $k-2$ and $k$ satisfies $k \le N-1$. We conclude that there is never a solution for $i=1$. Therefore there is never a degenerate operator associated to $H_1$.
\item If $k=1$ then $n$ and $m$ don't exist. Conclusion, in order to get degenerate operators we should have $N \ge 3$ which means that degenerate operators will pop up in frames with at least $3$ gauge groups (which correspond to $k=2$).
\end{itemize}
We can ask the more precise question: What is the first frame, $\CT_{k_{min}}$, when some operators degenerate? In order to answer that we have to try to maximize the l.h.s of \eqref{DegOpConstraint} and minimize the r.h.s. Therefore it is enough to look at $n=k_{min}-1$, $m=1$ and $i=k_{min}$ to determine $k_{min}$ (it could also have degenerate operators in $\CT_{k_{min}}$ not associated to $n=k_{min}-1$, $m=1$ and $i=k_{min}$). We obtain
\be 
k_{min}-2 = N-k_{min}-1 \quad \Rightarrow \quad k_{min} = \left\lceil \frac{N+1}{2} \right\rceil
\ee
So when we reach $\CT_{k_{min}}$ we start having this issue of degenerate operators.

Now let us solve \eqref{DegOpConstraint} in the case $k=N-1$. In this case, $1 \le m \le n \le N-2$ 
\begin{center}
\begin{tabular}{l c}
$i=2:$ & $n=N+m-3$ \\
& $\exists$ solution for $m=1$ \\
$i=3:$ & $n=N+m-4$ \\
& $\exists$ solution for $m=1,2$ \\
\quad $\vdots$ & \\
$i=N-1$: & $n=m$ \\
& $\exists$ solution for $m=1, \dots, N-2$
\end{tabular}
\end{center}
Conclusion, the degenerate operators (with respect to $V_i B_i \dots B_{N-2} C_{N-1} F_N$) that potentially coupled to $H_i$ in $\CT_{N-1}$ are:
\begin{center}
\begin{tabular}{l c c}  
$H_2:$ & $V_1 B_1 \dots B_{N-3} R_{N-2}$ & 1 operator \\
$H_3:$ & $V_1 B_1 \dots B_{N-4} R_{N-3}, \, V_2 B_2 \dots B_{N-3} R_{N-2}$ & 2 operators \\
$\quad \vdots$ & & \\
$H_{N-1}:$ & $V_1 R_1, \, V_2 R_2, \, \dots, \, V_{N-2} R_{N-2}$ & N-2 operators 
\end{tabular}
\end{center}
Finally, we can study the final frame $\CT_{Dec}$. The R-charges are the following
\begin{table}[H]
\centering
\begin{tabular}{|c|c|}
\hline
Fields & R-charges in $\CT_{Dec}$ \\
\hline
$v_i$ & $5 - 2F + (N - \frac{5}{2} + \frac{3i}{2})R_A + (2F-1)R_Q$ \\
\hline
$l_i$ & $2F-4 - (N-2+i)R_A - (2F-2)R_Q$ \\
\hline
$m_i$ & $2R_Q + (i-1)R_A$ \\
\hline
$h_j$ & $(N-i+1)R_A$ \\
\hline
$r_j$ & $2F-3 - (N + \frac{3j}{2} - \frac{1}{2})R_A - (2F-1)R_Q$ \\
\hline
$b_j$ & $\frac{1}{2}R_A$ \\
\hline
$a_j$ & $2 - R_A$ \\
\hline
$q_N$ & $1 - R_Q - \frac{1}{2}(N-1)R_A$ \\
\hline
\end{tabular}
\caption{R-charges in $\CT_{Dec}$ with $i=1, \dots, N$ and $j=1, \dots, N-1$.}
\label{table:2}
\end{table}

Now using the previous result for $\CT_{N-1}$, we can summarize all the candidates to be coupled to $h_i$ in the final frame $\CT_{Dec}$
\begin{center}
\begin{tabular}{l c c} 
$h_1:$ & $v_1 b_1 \dots b_{N-2} r_{N-1}$ & No degenerate operator \\ 
$h_2:$ & $v_1 b_1 \dots b_{N-3} r_{N-2}, \, v_2 b_2 \dots b_{N-2} r_{N-1}$ & 2 operators \\
$h_3:$ & $v_1 b_1 \dots b_{N-4} r_{N-3}, \, v_2 b_2 \dots b_{N-3} r_{N-2}, \, v_3 b_3 \dots B_{N-2} r_{N-1}$ & 3 operators \\
$\quad \vdots$ & & \\
$h_{N-1}:$ & $v_1 r_1, \, v_2 r_2, \, \dots, \, v_{N-2} r_{N-2}, \, v_{N-1} r_{N-1}$ & N-1 operators 
\end{tabular}
\end{center}

Now the question is obvious, which operator is the correct one? 

\noindent In the special case of $F=4$, we could use the same argument that we used in Section~\ref{USp6}. It goes as follows. When we reach the frame $\CT_{k_{min}}$ some operators become degenerate. In order to decide the correct operator, we start confining from the left (it is possible because in the case of $F=4$ the gauge group becomes $Usp(2)$). Then at some point we will discover that using the F-term equation for $H_1$ (which is never associated to a degenerate operator as we saw) we can select the correct operator associated to some $H_a$ (as in \eqref{SwitchingOp}). Then, the procedure is iterative meaning that we should re-use our previous results for $H_a$ and do more and more reconfinement to select all the correct operators associated to the other $H_b$. All in all, we end up in the frame $\CT_{N-1}$ with the following superpotential term
\be \label{ModifiedW1}
\delta \cW_{N-1} = H_1 V_1 B_1 \dots B_{N-2} C_{N-1} P_N + \sum_{i=2}^{N-1} H_i V_1 B_1 \dots B_{N-1-i} R_{N-i} 
\ee  
which becomes in the final frame $\CT_{Dec}$
\be \label{ModifiedW2}
\delta \cW_{Dec} = \sum_{i=1}^{N-1} h_i v_1 b_1 \dots b_{N-1-i} r_{N-i}
\ee  

Unfortunately, in the case of $F>4$ the previous argument fails because we cannot reconfine from the left. In this case we can deconfine the antisymmetric field $A_1$ but we didn't manage to find constraints and remove the degeneracy. Therefore, in this case the superpotential that we wrote in \eqref{UspTN-1} and \eqref{UspTDec} are ambiguous. We wrote them with the results \eqref{ModifiedW1}, \eqref{ModifiedW2} obtain in the case $F=4$ but it is logically possible that they are wrong for $F>4$.

\bibliographystyle{ytphys}
\bibliography{refs}
\end{document}